\documentclass[twocolumn]{aastex6}
\usepackage{amsmath}
\usepackage{rotating}
\usepackage{multirow}
\usepackage{setspace}
\usepackage{graphicx}

\slugcomment{DRAFT}

\shorttitle{SOMA Survey: High Luminosity Protostars}
\shortauthors{Liu et al.}

\begin{document}

\title{The SOFIA Massive (SOMA) Star Formation Survey. II.\\
High Luminosity Protostars}

\author{Mengyao Liu}
\affil{Dept. of Astronomy, University of Virginia, Charlottesville, Virginia 22904, USA}
\author{Jonathan C. Tan}
\affil{Dept. of Space, Earth \& Environment, Chalmers University of Technology, Gothenburg, Sweden and\\ Dept. of Astronomy, University of Virginia, Charlottesville, Virginia 22904, USA}
\author{James M. De Buizer}
\affil{SOFIA-USRA, NASA Ames Research Center, MS 232-12, Moffett Field, CA 94035, USA}
\author{Yichen Zhang}
\affil{Star and Planet Formation Laboratory, RIKEN Cluster for Pioneering Research, Wako, Saitama 351-0198, Japan}
\author{Maria T. Beltr\'{a}n}
\affil{INAF-Osservatorio Astrofisico di Arcetri, Largo E. Fermi 5, I-50125 Firenze, Italy}
\author{Jan E. Staff} 
\affil{College of Science and Math, University of Virgin Islands, St. Thomas, United States Virgin Islands 00802}
\author{Kei E. I. Tanaka}
\affil{Department of Earth and Space Science, Osaka University, Toyonaka, Osaka 560-0043, Japan and\\  ALMA Project, National Astronomical Observatory of Japan, Mitaka, Tokyo 181-8588, Japan}
\author{Barbara Whitney}
\affil{Department of Astronomy, University of Wisconsin-Madison, 475 N. Charter St, Madison, WI 53706, USA}
\author{Viviana Rosero}
\affil{National Radio Astronomy Observatory, 1003 Lopezville Rd., Socorro, NM 87801, USA}

\begin{abstract}
We present multi-wavelength images observed with {\it SOFIA}-FORCAST
from $\sim$10 to 40\,$\mu$m of seven high luminosity massive
protostars, as part of the {\it SOFIA} Massive (SOMA) Star Formation
Survey. Source morphologies at these wavelengths appear to be
influenced by outflow cavities and extinction from dense gas
surrounding the protostars.
%consistent with expectations from the Turbulent Core Accretion (TCA) models. 
Using these images, we build spectral energy distributions (SEDs) of
the protostars, also including archival data from {\it Spitzer}, {\it
  Herschel} and other facilities. Radiative transfer (RT) models of
Zhang \& Tan (2018), based on Turbulent Core Accretion theory,
%including disk wind outflow cavities, 
are then fit to the SEDs to estimate key properties of the
protostars. Considering the best five models fit to each source, the
protostars have masses $m_{*} \sim 12-64 \: M_{\odot}$ accreting at
rates of $\dot{m}_{*} \sim 10^{-4}-10^{-3} \: M_{\odot} \: \rm
yr^{-1}$ inside cores of initial masses $M_{c} \sim 100-500 \:
M_{\odot}$ embedded in clumps with mass surface densities $\Sigma_{\rm
  cl} \sim 0.1-3 \: \rm g \: cm^{-2}$ and span a luminosity range of
$10^{4} -10^{6} \: L_{\odot}$. Compared with the first eight
protostars in Paper I, the sources analyzed here are more luminous,
and thus likely to be more massive protostars. They are often in a
clustered environment or have a companion protostar relatively nearby.
From the range of parameter space of the models, we do not see any
evidence that $\Sigma_{\rm cl}$ needs to be high to form these massive
stars. For most sources the RT models provide reasonable fits to the
SEDs, though the cold clump material often influences the long
wavelength fitting.  However, for sources in very clustered
environments, the model SEDs may not be such a good description of the
data, indicating potential limitations of the models for these
regions.

\end{abstract}


\section{Introduction}

Massive stars play a key role in the regulation of galaxy environments
and their overall evolution, yet there is no consensus on their
formation mechanism. Theories range from Core Accretion (e.g.,
McLaughlin \& Pudritz 1996; McKee \& Tan 2003 [MT03]) in which massive
stars form via a monolithic collapse of a massive core, to Competitive
Accretion (e.g., Bonnell et al. 2001; Wang et al. 2010) in which
massive stars have most of the mass reservoir joining later and form
hand in hand with the formation of a cluster of mostly low-mass stars,
to Protostellar Collisions (Bonnell et al. 1998). The confusion
remains partly due to the difficulty of observations towards massive
star formation given the typically large distances and high extinction
of the regions.

Outflows appear to be a ubiquitous phenomenon in the formation of
stars of all masses. They may limit the formation efficiency from a
core since they expel material along polar directions. The resulting
outflow cavities have been proposed to affect the appearance of
massive sources in the mid-IR (MIR) up to $\sim$40 $\mu$m (De Buizer
2006; Zhang et al. 2013a) and this is seen in radiative transfer (RT)
calculations of the Turbulent Core Model of MT03 (e.g., Zhang et
al. 2013b, Zhang et al. 2014; Zhang \& Tan 2018).

Motivated by the need of observations of a larger sample of massive
protostars to test theoretical models of massive star formation, we
are carrying out the \textit{SOFIA} Massive (SOMA) Star Formation
Survey (PI: Tan). The overall goal is to obtain $\sim$ 10 to
40\,$\mu$m images with the \textit{SOFIA}-FORCAST instrument of a sample
of $\ga$ 50 high- and intermediate-mass protostars over a range of
evolutionary stages and environments, and then compare the observed
spectral energy distributions (SEDs) and image intensity profiles with
theoretical models.
%So far 24 sources have been observed.
The results and SED analysis of the first 8 sources of the survey have
been published by De Buizer et al. (2017) (hereafter Paper I).

%This work is our second paper of the series of the SOMA survey.
In this paper, we now present the next seven most luminous protostars
from the sample of completed observations, which are expected to be the
highest-mass protostars. 
% Yichen
In this work we still focus on the SED analysis. Comparison with the
image intensity profiles will be presented in a future paper.
The observations and data used are described in \S2. The analysis
methods are described in \S3. We present the MIR imaging and SED
fitting results in \S4 and discuss these results and their
implications in \S5. A summary is given in \S6.

\section{Observations}

\subsection{SOFIA Data}

%jctfinal - check this
The following seven sources, listed in order of decreasing isotropic
bolometric luminosity, were observed by
\textit{SOFIA}\footnote{\textit{SOFIA} is jointly operated by the
  Universities Space Research Association, Inc. (USRA), under NASA
  contract NAS2-97001, and the Deutsches SOFIA Institute (DSI) under
  DLR contract 50 OK 0901 to the University of Stuttgart.}  (Young et
al. 2012) with the FORCAST instrument (Herter et al. 2013) (see
Table~\ref{tab:sofia_obs}): G45.12+0.13; G309.92+0.48; G35.58-0.03;
IRAS 16562-3959; G305.20+0.21; G49.27-0.34; G339.88-1.26.

%Data were taken on Cycle 4 \textit{SOFIA} observing cycles. All observations were taken at an altitude between 39000 and 43000 ft, which typically yields precipitable water vapor overburdens of less than 25\,$\mu$m. All data were taken by employing the standard chop-nod observing technique used in the thermal infrared, with chop and nod throws sufficiently large to sample clear off-source sky.

\textit{SOFIA} data were calibrated by the \textit{SOFIA} pipeline
with a system of stellar calibrators taken across all flights in a
flight series and applied to all targets within that flight series
(see also the FORCAST calibration paper by Herter et
al. 2013). Corrections were also made for the airmass of the
sources. The main uncertainty in the \textit{SOFIA} calibrations is
caused by the apparent variability in the flux of the standard stars
throughout the flight and from flight to flight due to changing
atmospheric conditions. The calibration error is estimated to be in
the range $\sim$ 3\% - 7\%.

\begin{deluxetable*}{ccccccccc}
%\tabletypesize{\scriptsize}
\tablecaption{{\it SOFIA} FORCAST Observations: Observation Dates \& Exposure Times (seconds)\label{tab:sofia_obs}}
\tablewidth{14pt}
\tablehead{
\colhead{Source} & \colhead{R.A.(J2000)} & \colhead{Decl.(J2000)} & \colhead{$d$ (kpc)} & \colhead{Obs. Date} & \colhead{7.7$\:{\rm \mu m}$} & \colhead{19.7$\:{\rm \mu m}$}  & \colhead{31.5$\:{\rm \mu m}$} & \colhead{37.1$\:{\rm \mu m}$}
}
\startdata
G45.12+0.13 & 19$^h$13$^m$27$\fs$859 & $+$10$\arcdeg$53$\arcmin$36$\farcs$645 & 7.4 & 2016 Sep 17 & 2443 & 882 & 623 & 1387  \\
G309.92+0.48 & 13$^h$50$^m$41$\fs$847 & $-$61$\arcdeg$35$\arcmin$10$\farcs$40 & 5.5 & 2016 Jul 14 & 291 & 828 & 532 & 1691 \\
G35.58-0.03 & 18$^h$56$^m$22$\fs$563 & $+$02$\arcdeg$20$\arcmin$27$\farcs$660 & 10.2 & 2016 Sep 20 & 335 & 878 & 557 & 1484 \\
IRAS 16562-3959 & 16$^h$59$^m$41$\fs$63 & $-$40$\arcdeg$03$\arcmin$43$\farcs$61& 1.7 & 2016 Jul 17 & 1461 & 772 & 502 & 1243 \\
G305.20+0.21 & 13$^h$11$^m$10$\fs$49 & $-$62$\arcdeg$34$\arcmin$38$\farcs$8 & 4.1 & 2016 Jul 18 & 1671 & 763 & 539 & 1028 \\
G49.27-0.34 & 19$^h$23$^m$06$\fs$61 & $+$14$\arcdeg$20$\arcmin$12$\farcs$0 & 5.55 & 2016 Sep 20 & 290 & 716 & 664 & 1307 \\
G339.88-1.26 & 16$^h$52$^m$04$\fs$67 & $-$46$\arcdeg$08$\arcmin$34$\farcs$16 & 2.1 & 2016 Jul 20 & 1668 & 830 & 527 & 1383 \\
\enddata
\tablecomments{
The source positions listed here are the same as the positions of the
black crosses denoting the radio continuum peak (methanol maser in
G305.20) in each source in Figures 1, 2, 4, 6, 7, 9, 10. The ordering
of the sources is based on their isotropic luminosity estimate from
high to low (top to bottom). Source distances are from the literature,
discussed below.}
\end{deluxetable*}

\subsection{Other IR Data}

For all objects, data were retrieved from the \textit{Spitzer}
Heritage Archive from all four IRAC (Fazio et al. 2004) channels (3.6,
4.5, 5.8 and 8.0\,$\mu$m). In some cases, the sources are so bright
that they are saturated in the IRAC images and so these could not be
used to derive accurate fluxes. For IRAS 16562, we used unsaturated
\textit{WISE} archival data (3.4\,$\mu$m and 4.6\,$\mu$m) as a
substitute.

We also incorporated publicly-available imaging observations performed
with the \textit{Herschel Space
  Observatory}\footnote{\textit{Herschel} is an ESA space observatory
  with science instruments provided by European-led Principal
  Investigator consortia and with important participation from
  NASA. The \textit{Herschel} data used in this paper are taken from
  the Level 2 (flux-calibrated) images provided by the
  \textit{Herschel} Science Center via the NASA/IPAC Infrared Science
  Archive (IRSA), which is operated by the Jet Propulsion Laboratory,
  California Institute of Technology, under contract with NASA.}
(Pilbratt et al. 2010) and its PACS (Poglitsch et al. 2010) and SPIRE
(Griffin et al. 2010) instruments at 70, 160, 250, 350 and
500\,$\mu$m.

In addition to using these data for deriving multi-wavelength flux
densities of our sources, the \textit{Spitzer} 8\,$\mu$m and
\textit{Herschel} 70\,$\mu$m images are presented for comparison with
our \textit{SOFIA} images in \S\ref{S:indiv}. We note that the data
being analyzed here were typically collected within a time frame of
about 10 years (i.e., for the \textit{Spitzer}, \textit{Herschel}, and
\textit{SOFIA} observations).

We also present previously unpublished \textit{Gemini} 8-m data taken
with the instrument T-ReCS (De Buizer \& Fisher 2004) for sources
G309.92, G35.58, and G305.20. For both G309.92 and G35.58, only
11.7\,$\mu$m data were taken, with on-source exposures times of 304s
and 360s, respectively. For G305.20, we have images through ten T-ReCS
filters from 3.8\,$\mu$m (L-band) to 24.5\,$\mu$m, all with an
exposure time of 130s. Most T-ReCS filters have modest flux
calibration errors (for MIR observations) with standard
deviations between 2 and 10\%. For instance, the 11.7\,$\mu$m filter
has a 1-sigma flux calibration error of 3\%. Flux calibration through
certain filters, however, is more difficult due to the presence of
various atmospheric absorption lines contaminating the filter
bandpass, some of which can be highly variable. Those filters most
affected are the 7.7\,$\mu$m (21\%), 12.3 µm (19\%), 18.3 µm (15\%),
and 24.6 µm (23\%) filters (De Buizer et al. 2005).

NIR images from the VISTA/VVV3 (Minniti et al. 2010) and the
WFCAM/UKIDSS (Lawrence et al. 2007) surveys are also used to
investigate the environments of the protostellar sources and look for
association with the MIR counterparts.

\subsection{Astrometry}

%\textit{SOFIA} observations were performed in such a way using the simultaneous observations with the dichroic that the relative astrometry between the four \textit{SOFIA} images has been determined to be better than a FORCAST pixel ($\sim$0.77$\arcsec$). 
The absolute astrometry of the \textit{SOFIA} data comes from matching
the centroids of point sources in the \textit{SOFIA}
7\,$\mu$m image with the \textit{Spitzer} 8\,$\mu$m image (or shorter
IRAC wavelength, if saturated at 8\,$\mu$m). The relative astrometry
between the four \textit{SOFIA} images is reduced to be better than
0.4\arcsec, which is around half a FORCAST pixel. Thus the astrometry
precision is about 0.1\arcsec for the \textit{SOFIA} 7\,$\mu$m image
and 0.4\arcsec for longer wavelength \textit{SOFIA} images. The
\textit{Herschel} data can also be off in their absolute astrometry by
up to 5$\arcsec$. For all targets in this survey, we were able to find
point sources in common between the \textit{Herschel} image and
sources found in the \textit{SOFIA} or \textit{Spitzer} field of view
that allowed us to correct the \textit{Herschel} absolute
astrometry. The astrometry is then assumed to have errors of less than
1$\arcsec$.

The \textit{Gemini} images are calibrated using the \textit{Spitzer}
data and the astrometry precision is better than $\sim$
0.2\arcsec. The archival WISE data and NIR data from the VVV survey
and the UKIDSS survey were calibrated using 2MASS point source catalog
and should have a positional accuracy $<$ 0.1\arcsec. 

\section{Methods}

\subsection{SED Construction}\label{S:SED construction}

We follow the methods in Paper I and use PHOTUTILS, a Python package,
to measure the flux photometry. When building the SEDs, we try two
different methods. One is using fixed aperture size for all
wavelengths, which is our fiducial case. The aperture size is mainly
based on the \textit{Herschel} 70 $\mu$m image, which is typically
close to the peak of the SED, in order to capture the most flux from
the source, while minimizing contamination from other sources. We
assume this is the ``core'' scale from which the protostar forms as
described in the Turbulent Core Model (MT03).
%jctfinal - not sure about this
%Yichen
%The aperture radius is usually set at the conjunction of source
%emission and the background from the radial profile of the source.
%%
If there is no \textit{Herschel} data available, we use the
\textit{SOFIA} 37 $\mu$m image to determine the aperture
size. Sometimes we see multiple IR peaks in the aperture at shorter
wavelengths, but without corresponding resolved structures at longer
wavelengths, as in G45.12, G309.92, G35.58, and G49.27. This is a
combined effect of larger beam sizes at the longer wavelengths and the
fact that the emission from the secondary sources appears to be weaker
at longer wavelengths. Note that due to the limited size of the field
of view of the Gemini images, even for the fixed aperture method, we
adopt an aperture radius of 9\arcsec, 9\arcsec, and 10\arcsec \ for the 
photometry of the Gemini images of G309.92, G35.58 and G305.20,
respectively, which are the largest aperture sizes possible to allow
for background subtraction in each image.

The alternate method is to use variable aperture sizes for each
wavelength $<70\:\mu$m. In this case, we typically use smaller
apertures at shorter wavelength to exclude secondary sources that
appear resolved from the main massive protostar in the fiducial
aperture in the \textit{Spitzer} and \textit{SOFIA} images and compare
the effects on the SEDs. The aperture is always centered at the radio
continuum source (or the location of the methanol maser if there is no
radio emission as in G305.20), where we assume the protostar is
located.

After measuring the flux inside the aperture, we carry out background
subtraction using the median flux density in an annular region
extending from 1 to 2 aperture radii, as in Paper I, to remove general
background and foreground contamination and the effect of a cooler,
more massive clump surrounding the core at long wavelengths. The
aperture radii are typically several times larger than the beam sizes
for wavelengths $\leq$ 70 $\mu$m (and by greater factors for the fixed
aperture method that uses the 70 $\mu$m aperture radii across all
bands). At wavelengths $>$ 70 $\mu$m, the fixed aperture radius set at
70 $\mu$m is always used, and the aperture diameter is still usually
larger than the image resolution (except for G305.20 whose fixed
aperture diameter becomes similar to the resolution at the longest
wavelength 500 $\mu$m).

\subsection{Zhang \& Tan Radiative Transfer Models}\label{S: ZT models}

We use Zhang \& Tan (2018, [ZT18]) radiative transfer (RT) models
(hereafter ZT models) to fit the SEDs and derive key physical
parameters of the protostars. In a series of papers, Zhang \& Tan
(2011), Zhang et al. (2013b), Zhang et al. (2014) and ZT18 have
developed models for the evolution of high- and intermediate-mass
protostars based on the Turbulent Core Model (MT03). In this model,
massive stars are formed from pre-assembled massive pre-stellar cores,
supported by internal pressure that is provided by a combination of
turbulence and magnetic fields. With various analytic or semi-analytic
solutions, they calculate the properties of a protostellar core with
different components, including the protostar, disk, infall envelope,
outflow, and their evolutions, self-consistently from given initial
conditions. The main free parameters in this model grid are: the
initial mass of the core $M_{c}$; the mass surface density of the
clump that the core is embedded in $\Sigma_{\rm cl}$; and the
protostellar mass, $m_{*}$, which indicates the evolutionary stage.
In addition, there are secondary parameters of inclination angle of
line of sight to the outflow axis, $\theta_{\rm view}$, and the level
of foreground extinction, $A_{V}$.

The evolutionary history of a protostar from a given set of initial
conditions ($M_{c}$ and $\Sigma_{\rm cl}$) is referred to as an
evolutionary track, and a particular moment on such a track is a
specified $m_{*}$. Therefore the model grid is of three dimensions
($M_{c}$-$\Sigma_{\rm cl}$-$m_{*}$), including the entire set of
tracks. Currently, $M_{c}$ is sampled at 10, 20, 30, 40, 50, 60, 80,
100, 120, 160, 200, 240, 320, 400, 480 $M_{\odot}$, $\Sigma_{\rm cl}$
is sampled at 0.1, 0.32, 1, 3.2 g $\rm cm^{-2}$, forming 60
evolutionary tracks. Then $m_{*}$ is sampled at 0.5, 1, 2, 4, 8, 12,
16, 24, 32, 48, 64, 96, 128, 160 $M_{\odot}$. Note that not all of
these $m_{*}$ are sampled for each track. In particular, the maximum
protostellar mass is limited by the final stellar mass achieved in a
given evolutionary track. As a result, there are 432 different
physical models defined by different sets of $M_{c}$, $\Sigma_{\rm
  cl}$ and $m_{*}$.

There are several things to note about the models. First, the models
describe one protostar forming through monolithic collapse from the
parent core. The formation of binary and multiple systems is not
included in the models. Second, compared with the Robitaille et
al. (2007) RT models that mostly focus on lower-mass protostars, the
ZT18 model grid has broader parameter space relevant to high pressure,
high density and thus high accretion rate conditions of massive star
formation, while keeping the number of free parameters low.
%, which allows for the high accretion rate, high
%density cores, etc, in high- and intermediate-mass start formation (De
%Buizer et al. 2017). Even though focusing on high-mass star formation,
%the models can also be extended to low-mass star formation (Zhang \&
%Tan 2015).
Third, the models do not explicitly include the clump component, which
contributes to foreground extinction at short wavelengths and
additional emission at long wavelengths. The former effect is
compensated for by the free parameter $A_{V}$. The latter effect
requires the model grid fitting to be done on
clump-envelope-background-subtracted SEDs. Fourth, the aperture scale
for the measured SED is not considered in the fitting process. The
predicted SEDs in the model grid are total SEDs, which include modest contributions from parts of the outflow that extend beyond
the core. We assume with the aperture adopted we also measure the
total emission from the protostar and ideally the models that describe
that observed SED best would predict a similar scale (this can be
checked after the fitting results are returned).  Fifth, PAH
emission and thermal emission from transiently (single-photon) heated
very small grains at $\la$ 8 $\mu$m is not modeled, and so our method
is to use the SEDs at these wavelengths as upper limits. Lastly, while
the general trends of the features of the SEDs are determined by the
initial/environmental conditions and evolution, some detailed
features, such as the peak wavelength and long-wavelength spectral
index, may be affected by the particular dust models used in the
radiative transfer simulations.

\subsection{SED Fitting}\label{S:SED fitting}

When fitting the SEDs to the models, we use our fiducial case, i.e.,
using fixed aperture size for all wavelengths, and set data points at
wavelengths $\leq$ 8 $\mu$m as upper limits since the effects of PAH
emission and thermal emission from very small grains are not included
in the ZT RT models. For G309, the Spitzer 4.5 $\mu$m, 5.8 $\mu$m and
8 $\mu$m data have a ghosting problem. For G45.12 and IRAS~16562 all
Spitzer data have ghosting problems. Thus we do not use these data for
the SED fitting. The error bars are set to be the larger of either
10\% of the clump background-subtracted flux density to account
for calibration error, or the value of the estimated clump
background flux density (see \S\ref{S:SED construction}),
which is used for background subtraction, given that order unity
fluctuations in the surrounding background flux are often seen.

The fitting procedure involves convolving model SEDs with the filter
response functions for the various telescope bands. Source distances
are adopted from the literature. For each source, we present the five
best-fitting models. Again we note that the SED model fitting performed
here assumes that there is a single dominant source of luminosity,
i.e., effects of multiple sources, including unresolved binaries, are
not accounted for. This is a general limitation and caveat associated
with this method as discussed in Paper I.
%Finally, the models used in this paper assume smoothly varying or
%constant accretion rates.

\section{Results}

The types of multi-wavelength data available for each source, the flux
densities derived, and the aperture sizes adopted are listed in
Table~\ref{tab:flux}. $F_{\lambda, \rm fix}$ is the flux density
derived with a fixed aperture size and $F_{\lambda, \rm var}$ is the
flux density derived with a variable aperture size. The value of flux
density listed in the upper row of each source is derived with
background subtraction, while that derived without background
subtraction is listed in brackets in the lower row. The \textit{SOFIA}
images for each source are presented in \S4.1. General results of the
\textit{SOFIA} imaging are summarized in \S4.2. The SEDs and fitting
results are presented in \S4.3.

%\clearpage
\begin{sidewaystable*}% Landscape page
\centering
\setlength{\tabcolsep}{0pt}
\renewcommand{\arraystretch}{0.8}
\vspace{-3in}
\begin{deluxetable*}{cc|ccc|ccc|ccc|ccc|ccc|ccc|ccc}
\tabletypesize{\scriptsize}
\tablecaption{Integrated Flux Densities\label{tab:flux}}
\tablewidth{18pt}
\tablehead{
\colhead{Facility} &\colhead{$\lambda$} &\colhead{F$_{\lambda,\rm fix}$}\tablenotemark{a} &\colhead{F$_{\lambda,\rm var}$} \tablenotemark{b} & \colhead{R$_{\rm ap}$} \tablenotemark{c} &\colhead{F$_{\lambda,\rm fix}$} &\colhead{F$_{\lambda,\rm var}$} & \colhead{R$_{\rm ap}$}&\colhead{F$_{\lambda,\rm fix}$} &\colhead{F$_{\lambda,\rm var}$} & \colhead{R$_{\rm ap}$}&\colhead{F$_{\lambda,\rm fix}$} &\colhead{F$_{\lambda,\rm var}$} & \colhead{R$_{\rm ap}$}&\colhead{F$_{\lambda,\rm fix}$} &\colhead{F$_{\lambda,\rm var}$} & \colhead{R$_{\rm ap}$}&\colhead{F$_{\lambda,\rm fix}$} &\colhead{F$_{\lambda,\rm var}$} & \colhead{R$_{\rm ap}$}&\colhead{F$_{\lambda,\rm fix}$} &\colhead{F$_{\lambda,\rm var}$} & \colhead{R$_{\rm ap}$} \\
\colhead{} &\colhead{($\mu$m)} &\colhead{(Jy)} &\colhead{(Jy)} & \colhead{($\arcsec$)} &\colhead{(Jy)} &\colhead{(Jy)} & \colhead{($\arcsec$)} &\colhead{(Jy)} &\colhead{(Jy)} & \colhead{($\arcsec$)} &\colhead{(Jy)} &\colhead{(Jy)} & \colhead{($\arcsec$)} &\colhead{(Jy)} &\colhead{(Jy)} & \colhead{($\arcsec$)} &\colhead{(Jy)} &\colhead{(Jy)} & \colhead{($\arcsec$)} &\colhead{(Jy)} &\colhead{(Jy)} & \colhead{($\arcsec$)} 
}
\startdata
 &  & \multicolumn{3}{c}{G45.12+0.13} &  \multicolumn{3}{c}{G309.92+0.48} &  \multicolumn{3}{c}{G35.58-0.03} &  \multicolumn{3}{c}{IRAS 16562}  &  \multicolumn{3}{c}{G305.20+0.21}  &  \multicolumn{3}{c}{G49.27-0.34} &  \multicolumn{3}{c}{G339.88-1.26}  \\
\hline
\multirow{2}{*}{WISE} & \multirow{2}{*}{3.4} & \multirow{2}{*}{...} & \multirow{2}{*}{...} & \multirow{2}{*}{...} & \multirow{2}{*}{...} & \multirow{2}{*}{...} & \multirow{2}{*}{...} & \multirow{2}{*}{...} & \multirow{2}{*}{...} & \multirow{2}{*}{...} & 1.62 & 0.62 & \multirow{2}{*}{12.0} & \multirow{2}{*}{...} & \multirow{2}{*}{...} & \multirow{2}{*}{...} & \multirow{2}{*}{...} & \multirow{2}{*}{...} & \multirow{2}{*}{...} & \multirow{2}{*}{...} & \multirow{2}{*}{...} & \multirow{2}{*}{...} \\
&   &   &   &   &   &   &   &   &   &   & (2.53) & (0.89) &   &   &   &   &   &   &   &   &   &   \\
\hline
\multirow{2}{*}{{\it Spitzer}/IRAC} & \multirow{2}{*}{3.6} & 5.04 & 2.83 & \multirow{2}{*}{12.0} & 2.68 & 1.56 & \multirow{2}{*}{6.0} & 0.51 & 0.05 & \multirow{2}{*}{6.0} & \multirow{2}{*}{...} & \multirow{2}{*}{...} & \multirow{2}{*}{...} & 1.15 & 0.62 & \multirow{2}{*}{4.6} & 0.10 & 0.05 & \multirow{2}{*}{7.7} & 1.10 & 0.39 & \multirow{2}{*}{12.0} \\
&   & (5.64) & (3.04) &   & (2.93) & (1.65) &   & (0.76) & (0.09) &   &   &   &   & (1.44) & (0.69) &   & (0.32) & (0.07) &   & (1.65) & (0.56) &   \\
\hline
\multirow{2}{*}{Gemini/T-ReCS} & \multirow{2}{*}{3.8} & \multirow{2}{*}{...} & \multirow{2}{*}{...} & \multirow{2}{*}{...} & \multirow{2}{*}{...} & \multirow{2}{*}{...} & \multirow{2}{*}{...} & \multirow{2}{*}{...} & \multirow{2}{*}{...} & \multirow{2}{*}{...} & \multirow{2}{*}{...} & \multirow{2}{*}{...} & \multirow{2}{*}{...} & 0.82 & 0.83 & \multirow{2}{*}{2.0} & \multirow{2}{*}{...} & \multirow{2}{*}{...} & \multirow{2}{*}{...} & \multirow{2}{*}{...} & \multirow{2}{*}{...} & \multirow{2}{*}{...} \\
&   &   &   &   &   &   &   &   &   &   &   &   &   & (1.10) & (0.87) &   &   &   &   &   &   &    \\
\hline
\multirow{2}{*}{{\it Spitzer}/IRAC} & \multirow{2}{*}{4.5} & 7.30 & 4.79 & \multirow{2}{*}{12.0} & 4.75 & 2.92 & \multirow{2}{*}{6.0} & 0.52 & 0.12 & \multirow{2}{*}{6.0} & \multirow{2}{*}{...} & \multirow{2}{*}{...} & \multirow{2}{*}{...} & 2.89 & 2.00 & \multirow{2}{*}{4.6} & 0.77 & 0.55 & \multirow{2}{*}{7.7} & 2.90 & 1.75 & \multirow{2}{*}{12.0} \\
&   & (7.87) & (5.08) &   & (4.98) & (3.12) &   & (0.73) & (0.17) &   &   &   &   & (3.18) & (2.11) &   & (0.99) & (0.59) &   & (3.67) & (2.01) &    \\
\hline
\multirow{2}{*}{WISE} & \multirow{2}{*}{4.6} & \multirow{2}{*}{...} & \multirow{2}{*}{...} & \multirow{2}{*}{...} & \multirow{2}{*}{...} & \multirow{2}{*}{...} & \multirow{2}{*}{...} & \multirow{2}{*}{...} & \multirow{2}{*}{...} & \multirow{2}{*}{...} & 4.20 & 0.73 & \multirow{2}{*}{12.0} & \multirow{2}{*}{...} & \multirow{2}{*}{...} & \multirow{2}{*}{...} & \multirow{2}{*}{...} & \multirow{2}{*}{...} & \multirow{2}{*}{...} & \multirow{2}{*}{...} & \multirow{2}{*}{...} & \multirow{2}{*}{...} \\
&   &   &   &   &   &   &   &   &   &   & (5.42) & (1.56) &   &   &   &   &   &   &   &   &   &    \\
\hline
\multirow{2}{*}{Gemini/T-ReCS} & \multirow{2}{*}{4.7} & \multirow{2}{*}{...} & \multirow{2}{*}{...} & \multirow{2}{*}{...} & \multirow{2}{*}{...} & \multirow{2}{*}{...} & \multirow{2}{*}{...} & \multirow{2}{*}{...} & \multirow{2}{*}{...} & \multirow{2}{*}{...} & \multirow{2}{*}{...} & \multirow{2}{*}{...} & \multirow{2}{*}{...} & 2.95 & 2.77 & \multirow{2}{*}{2.0} & \multirow{2}{*}{...} & \multirow{2}{*}{...} & \multirow{2}{*}{...} & \multirow{2}{*}{...} & \multirow{2}{*}{...} & \multirow{2}{*}{...} \\
&   &   &   &   &   &   &   &   &   &   &   &   &   & (3.35) & (2.85) &   &   &   &   &   &   &   \\
\hline
\multirow{2}{*}{{\it Spitzer}/IRAC} & \multirow{2}{*}{5.8} & 41.05 & 22.97 & \multirow{2}{*}{12.0} & 18.39 & 15.11 & \multirow{2}{*}{14.5} & 2.22 & 0.30 & \multirow{2}{*}{6.0} & \multirow{2}{*}{...} & \multirow{2}{*}{...} & \multirow{2}{*}{...} & 6.95 & 4.14 & \multirow{2}{*}{4.6} & 2.07 & 1.55 & \multirow{2}{*}{7.7} & \multirow{2}{*}{...} & \multirow{2}{*}{...} & \multirow{2}{*}{...}  \\
&   & (45.17) & (24.95) &   & (19.66) & (16.18) &   & (3.91) & (0.60) &   &   &   &   & (9.24) & (4.56) &   & (3.75) & (1.73) &   &   &   &    \\
\hline
\multirow{2}{*}{SOFIA/FORCAST} & \multirow{2}{*}{7.7} & 103.95 & 57.19 & \multirow{2}{*}{12.0} & 48.04 & 35.46 & \multirow{2}{*}{14.5} & 8.95 & 0.92 & \multirow{2}{*}{6.0} & 77.53 & 58.03 & \multirow{2}{*}{12.0} & 24.90 & 13.70 & \multirow{2}{*}{4.6} & 4.10 & 2.77 & \multirow{2}{*}{7.7} & 11.11 & 1.09 & \multirow{2}{*}{6.0} \\
&   & (92.89) & (60.84) &   & (47.17) & (37.55) &   & (5.83) & (1.62) &   & (81.72) & (60.63) &   & (26.31) & (14.95) &   & (4.20) & (3.02) &   & (17.69) & (1.67) &    \\
\hline
\multirow{2}{*}{Gemini/T-ReCS} & \multirow{2}{*}{7.9} & \multirow{2}{*}{...} & \multirow{2}{*}{...} & \multirow{2}{*}{...} & \multirow{2}{*}{...} & \multirow{2}{*}{...} & \multirow{2}{*}{...} & \multirow{2}{*}{...} & \multirow{2}{*}{...} & \multirow{2}{*}{...} & \multirow{2}{*}{...} & \multirow{2}{*}{...} & \multirow{2}{*}{...} & 9.39 & 12.05 & \multirow{2}{*}{2.0} & \multirow{2}{*}{...} & \multirow{2}{*}{...} & \multirow{2}{*}{...} & \multirow{2}{*}{...} & \multirow{2}{*}{...} & \multirow{2}{*}{...} \\
&   &   &   &   &   &   &   &   &   &   &   &   &   & (12.88) & (12.28) &   &   &   &   &   &   &   \\
\hline
\multirow{2}{*}{{\it Spitzer}/IRAC} & \multirow{2}{*}{8.0} & 72.80 & 25.68 & \multirow{2}{*}{12.0} & 28.21 & 20.29 & \multirow{2}{*}{14.5} & 5.15 & 0.73 & \multirow{2}{*}{6.0} & \multirow{2}{*}{...} & \multirow{2}{*}{...} & \multirow{2}{*}{...} & 17.92 & 8.62 & \multirow{2}{*}{4.6} & 2.34 & 1.78 & \multirow{2}{*}{7.7} & \multirow{2}{*}{...} & \multirow{2}{*}{...} & \multirow{2}{*}{...} \\
&   & (84.64) & (30.73) &   & (33.57) & (22.56) &   & (9.58) & (1.46) &   &   &   &   & (23.82) & (9.75) &   & (7.00) & (2.18) &   &   &   &    \\
\hline
\multirow{2}{*}{Gemini/T-ReCS} & \multirow{2}{*}{8.8} & \multirow{2}{*}{...} & \multirow{2}{*}{...} & \multirow{2}{*}{...} & \multirow{2}{*}{...} & \multirow{2}{*}{...} & \multirow{2}{*}{...} & \multirow{2}{*}{...} & \multirow{2}{*}{...} & \multirow{2}{*}{...} & \multirow{2}{*}{...} & \multirow{2}{*}{...} & \multirow{2}{*}{...} & 12.81 & 14.85 & \multirow{2}{*}{2.0} & \multirow{2}{*}{...} & \multirow{2}{*}{...} & \multirow{2}{*}{...} & \multirow{2}{*}{...} & \multirow{2}{*}{...} & \multirow{2}{*}{...} \\
&   &   &   &   &   &   &   &   &   &   &   &   &   & (15.58) & (15.01) &   &   &   &   &   &   &  \\
\hline
\multirow{2}{*}{Gemini/T-ReCS} & \multirow{2}{*}{9.7} & \multirow{2}{*}{...} & \multirow{2}{*}{...} & \multirow{2}{*}{...} & \multirow{2}{*}{...} & \multirow{2}{*}{...} & \multirow{2}{*}{...} & \multirow{2}{*}{...} & \multirow{2}{*}{...} & \multirow{2}{*}{...} & \multirow{2}{*}{...} & \multirow{2}{*}{...} & \multirow{2}{*}{...} & 17.14 & 17.24 & \multirow{2}{*}{2.0} & \multirow{2}{*}{...} & \multirow{2}{*}{...} & \multirow{2}{*}{...} & \multirow{2}{*}{...} & \multirow{2}{*}{...} & \multirow{2}{*}{...} \\
&   &   &   &   &   &   &   &   &   &   &   &   &   & (17.65) & (17.41) &   &   &   &   &   &   &  \\
\hline
\multirow{2}{*}{Gemini/T-ReCS} & \multirow{2}{*}{10.4} & \multirow{2}{*}{...} & \multirow{2}{*}{...} & \multirow{2}{*}{...} & \multirow{2}{*}{...} & \multirow{2}{*}{...} & \multirow{2}{*}{...} & \multirow{2}{*}{...} & \multirow{2}{*}{...} & \multirow{2}{*}{...} & \multirow{2}{*}{...} & \multirow{2}{*}{...} & \multirow{2}{*}{...} & 25.66 & 25.19 & \multirow{2}{*}{2.0} & \multirow{2}{*}{...} & \multirow{2}{*}{...} & \multirow{2}{*}{...} & \multirow{2}{*}{...} & \multirow{2}{*}{...} & \multirow{2}{*}{...}  \\
&   &   &   &   &   &   &   &   &   &   &   &   &   & (26.33) & (25.41) &   &   &   &   &   &   &   \\
\hline
\multirow{2}{*}{Gemini/T-ReCS} & \multirow{2}{*}{11.7} & \multirow{2}{*}{...} & \multirow{2}{*}{...} & \multirow{2}{*}{...} & 79.11 & 78.60 & \multirow{2}{*}{6.0} & 2.19 & 2.15 & \multirow{2}{*}{6.0} & \multirow{2}{*}{...} & \multirow{2}{*}{...} & \multirow{2}{*}{...} & 40.91 & 44.94 & \multirow{2}{*}{2.0} & \multirow{2}{*}{...} & \multirow{2}{*}{...} & \multirow{2}{*}{...} & \multirow{2}{*}{...} & \multirow{2}{*}{...} & \multirow{2}{*}{...} \\
&   &   &   &   & (80.37) & (79.26) &   & (2.25) & (2.21) &   &   &   &   & (47.11) & (45.33) &   &   &   &   &   &   &    \\
\hline
\multirow{2}{*}{Gemini/T-ReCS} & \multirow{2}{*}{12.3} & \multirow{2}{*}{...} & \multirow{2}{*}{...} & \multirow{2}{*}{...} & \multirow{2}{*}{...} & \multirow{2}{*}{...} & \multirow{2}{*}{...} & \multirow{2}{*}{...} & \multirow{2}{*}{...} & \multirow{2}{*}{...} & \multirow{2}{*}{...} & \multirow{2}{*}{...} & \multirow{2}{*}{...} & 54.23 & 56.18 & \multirow{2}{*}{2.0} & \multirow{2}{*}{...} & \multirow{2}{*}{...} & \multirow{2}{*}{...} & \multirow{2}{*}{...} & \multirow{2}{*}{...} & \multirow{2}{*}{...} \\
&   &   &   &   &   &   &   &   &   &   &   &   &   & (59.14) & (56.68) &   &   &   &   &   &   &  \\
\hline
\multirow{2}{*}{Gemini/T-ReCS} & \multirow{2}{*}{18.3} & \multirow{2}{*}{...} & \multirow{2}{*}{...} & \multirow{2}{*}{...} & \multirow{2}{*}{...} & \multirow{2}{*}{...} & \multirow{2}{*}{...} & \multirow{2}{*}{...} & \multirow{2}{*}{...} & \multirow{2}{*}{...} & \multirow{2}{*}{...} & \multirow{2}{*}{...} & \multirow{2}{*}{...} & 137 & 161 & \multirow{2}{*}{2.0} & \multirow{2}{*}{...} & \multirow{2}{*}{...} & \multirow{2}{*}{...} & \multirow{2}{*}{...} & \multirow{2}{*}{...} & \multirow{2}{*}{...}  \\
&   &   &   &   &   &   &   &   &   &   &   &   &   & (174) & (164) &   &   &   &   &   &   &    \\
\hline
\multirow{2}{*}{SOFIA/FORCAST} & \multirow{2}{*}{19.7} & 1128 & 976 & \multirow{2}{*}{12.0} & 380 & 345 & \multirow{2}{*}{10.0} & 21.78 & 18.40 & \multirow{2}{*}{7.0} & 254 & 212 & \multirow{2}{*}{12.0} & 282 & 194 & \multirow{2}{*}{4.6} & 2.97 & 2.12 & \multirow{2}{*}{11.0} & 26.94 & 24.87 & \multirow{2}{*}{6.0}  \\
&   & (1087) & (988) &   & (376) & (350) &   & (25.61) & (19.32) &   & (241) & (214) &   & (280) & (201) &   & (2.84) & (2.25) &   & (19.63) & (25.66) &    \\
\hline
\multirow{2}{*}{Gemini/T-ReCS} & \multirow{2}{*}{24.5} & \multirow{2}{*}{...} & \multirow{2}{*}{...} & \multirow{2}{*}{...} & \multirow{2}{*}{...} & \multirow{2}{*}{...} & \multirow{2}{*}{...} & \multirow{2}{*}{...} & \multirow{2}{*}{...} & \multirow{2}{*}{...} & \multirow{2}{*}{...} & \multirow{2}{*}{...} & \multirow{2}{*}{...} & 311 & 375 & \multirow{2}{*}{2.0} & \multirow{2}{*}{...} & \multirow{2}{*}{...} & \multirow{2}{*}{...} & \multirow{2}{*}{...} & \multirow{2}{*}{...} & \multirow{2}{*}{...}  \\
&   &   &   &   &   &   &   &   &   &   &   &   &   & (428) & (385) &   &   &   &   &   &   &   \\
\hline
\multirow{2}{*}{SOFIA/FORCAST} & \multirow{2}{*}{31.5} & 3077 & 2345 & \multirow{2}{*}{12.0} & 1896 & 1700 & \multirow{2}{*}{12.0} & 276 & 210 & \multirow{2}{*}{7.7} & 2078 & 1758 & \multirow{2}{*}{16.0} & 687 & 521 & \multirow{2}{*}{7.7} & 63.37 & 41.77 & \multirow{2}{*}{11.0} & 720 & 541 & \multirow{2}{*}{7.7} \\
&   & (3048) & (2423) &   & (1899) & (1735) &   & (275) & (221) &   & (2073) & (1797) &   & (710) & (546) &   & (69.24) & (45.96) &   & (714) & (566) &   \\
\hline
\multirow{2}{*}{SOFIA/FORCAST} & \multirow{2}{*}{37.1} & 4126 & 2952 & \multirow{2}{*}{12.0} & 2601 & 2298 & \multirow{2}{*}{12.0} & 525 & 365 & \multirow{2}{*}{7.7} & 3015 & 2444 & \multirow{2}{*}{16.0} & 892 & 654 & \multirow{2}{*}{7.7} & 89.39 & 58.83 & \multirow{2}{*}{11.0} & 1202 & 815 & \multirow{2}{*}{7.7} \\
&   & (4112) & (3082) &   & (2607) & (2352) &   & (531) & (394) &   & (3032) & (2531) &   & (925) & (694) &   & (86.18) & (62.34) &   & (1210) & (870) &   \\
\hline
\multirow{2}{*}{{\it Herschel}/PACS} & \multirow{2}{*}{70.0} & 5848 & 5848 & \multirow{2}{*}{48.0} & 3403 & 3403 & \multirow{2}{*}{32.0} & 1538 & 1538 & \multirow{2}{*}{25.6} & \multirow{2}{*}{...} & \multirow{2}{*}{...} & \multirow{2}{*}{...} & 1250 & 1250 & \multirow{2}{*}{16.0} & 449 & 449 & \multirow{2}{*}{28.8} & 3610 & 3610 & \multirow{2}{*}{32.0} \\
&   & (6205) & (6205) &   & (3536) & (3536) &   & (1647) & (1647) &   &   &   &   & (1617) & (1617) &   & (593) & (593) &   & (3846) & (3846) &     \\
\hline
\multirow{2}{*}{{\it Herschel}/PACS} & \multirow{2}{*}{160.0} & 3517 & 3517 & \multirow{2}{*}{48.0} & 2088 & 2088 & \multirow{2}{*}{32.0} & 968 & 968 & \multirow{2}{*}{25.6} & \multirow{2}{*}{...} & \multirow{2}{*}{...} & \multirow{2}{*}{...} & 644 & 644 & \multirow{2}{*}{16.0} & 864 & 864 & \multirow{2}{*}{28.8} & 2723 & 2723 & \multirow{2}{*}{32.0} \\
&   & (4045) & (4045) &   & (2454) & (2454) &   & (1193) & (1193) &   &   &   &   & (1032) & (1032) &   & (1198) & (1198) &   & (3046) & (3046) &    \\
\hline
\multirow{2}{*}{{\it Herschel}/SPIRE} & \multirow{2}{*}{250.0} & 1506 & 1506 & \multirow{2}{*}{48.0} & \multirow{2}{*}{...} & \multirow{2}{*}{...} & \multirow{2}{*}{...} & 404 & 404 & \multirow{2}{*}{25.6} & \multirow{2}{*}{...} & \multirow{2}{*}{...} & \multirow{2}{*}{...} & 234 & 234 & \multirow{2}{*}{16.0} & 517 & 517 & \multirow{2}{*}{28.8} & \multirow{2}{*}{...} & \multirow{2}{*}{...} & \multirow{2}{*}{...} \\
&   & (1796) & (1796) &   &   &   &   & (545) & (545) &   &   &   &   & (433) & (433) &   & (736) & (736) &   &   &   &     \\
\hline
\multirow{2}{*}{{\it Herschel}/SPIRE} & \multirow{2}{*}{350.0} & 469 & 469 & \multirow{2}{*}{48.0} & 289 & 289 & \multirow{2}{*}{32.0} & 129 & 129 & \multirow{2}{*}{25.6} & \multirow{2}{*}{...} & \multirow{2}{*}{...} & \multirow{2}{*}{...} & 60 & 60 & \multirow{2}{*}{16.0} & 193 & 193 & \multirow{2}{*}{28.8} & \multirow{2}{*}{...} & \multirow{2}{*}{...} & \multirow{2}{*}{...} \\
&   & (591) & (591) &   & (395) & (395) &   & (191) & (191) &   &   &   &   & (143) & (143) &   & (292) & (292) &   &   &   &   \\
\hline
\multirow{2}{*}{{\it Herschel}/SPIRE} & \multirow{2}{*}{500.0} & 136 & 136 & \multirow{2}{*}{48.0} & 80 & 80 & \multirow{2}{*}{32.0} & 30.87 & 30.87 & \multirow{2}{*}{25.6} & 267 & 267 & \multirow{2}{*}{32.0} & 20.61 & 20.61 & \multirow{2}{*}{16.0} & 54.19 & 54.19 & \multirow{2}{*}{28.8} & \multirow{2}{*}{...} & \multirow{2}{*}{...} & \multirow{2}{*}{...}  \\
&   & (187) & (187) &   & (122) & (122) &   & (56.16) & (56.16) &   & (374) & (374) &   & (59.45) & (59.45) &   & (94.92) & (94.92) &   &   &   &    \\
\enddata
\tablenotetext{a}{Flux density derived with a fixed aperture size of the 70\,$\mu$m data.}
\tablenotetext{b}{Flux density derived with various aperture sizes.}
\tablenotetext{c}{Aperture radius.}
\tablecomments{The value of flux density in the upper row is derived with background subtraction. The value in the bracket in the lower line is flux density derived without background subtraction. \\ F$_{\lambda,\rm fix}$ of the Gemini images of G309.92, G35.58 and G305.20 are derived with an aperture radius of 9\arcsec, 9\arcsec, and 10\arcsec, respectively. See more detail in \S\ref{S:SED construction}.}
\end{deluxetable*}
\end{sidewaystable*}

\subsection{Description of Individual Sources}\label{S:indiv}

In this section we describe the MIR morphology of each source and also
try to identify the nature of the structures revealed by our
\textit{SOFIA} or \textit{Gemini} imaging, together with archival NIR
data and other data from the literature.
%The sources are often in a clustered environment or have a companion
%protostar relatively nearby. Outflow cavities, dust extinction and
%potential disks seem to play an important role in the regulation of
%the MIR morphology, which is consistent with the expectations of the
%Core Accretion model. By investigating the detailed structure and
%context of the protostars, we illustrate the diversity and complexity
%of massive star forming regions.}

\subsubsection{G45.12+0.13}

\begin{figure*}
\epsscale{1.2}
\plotone{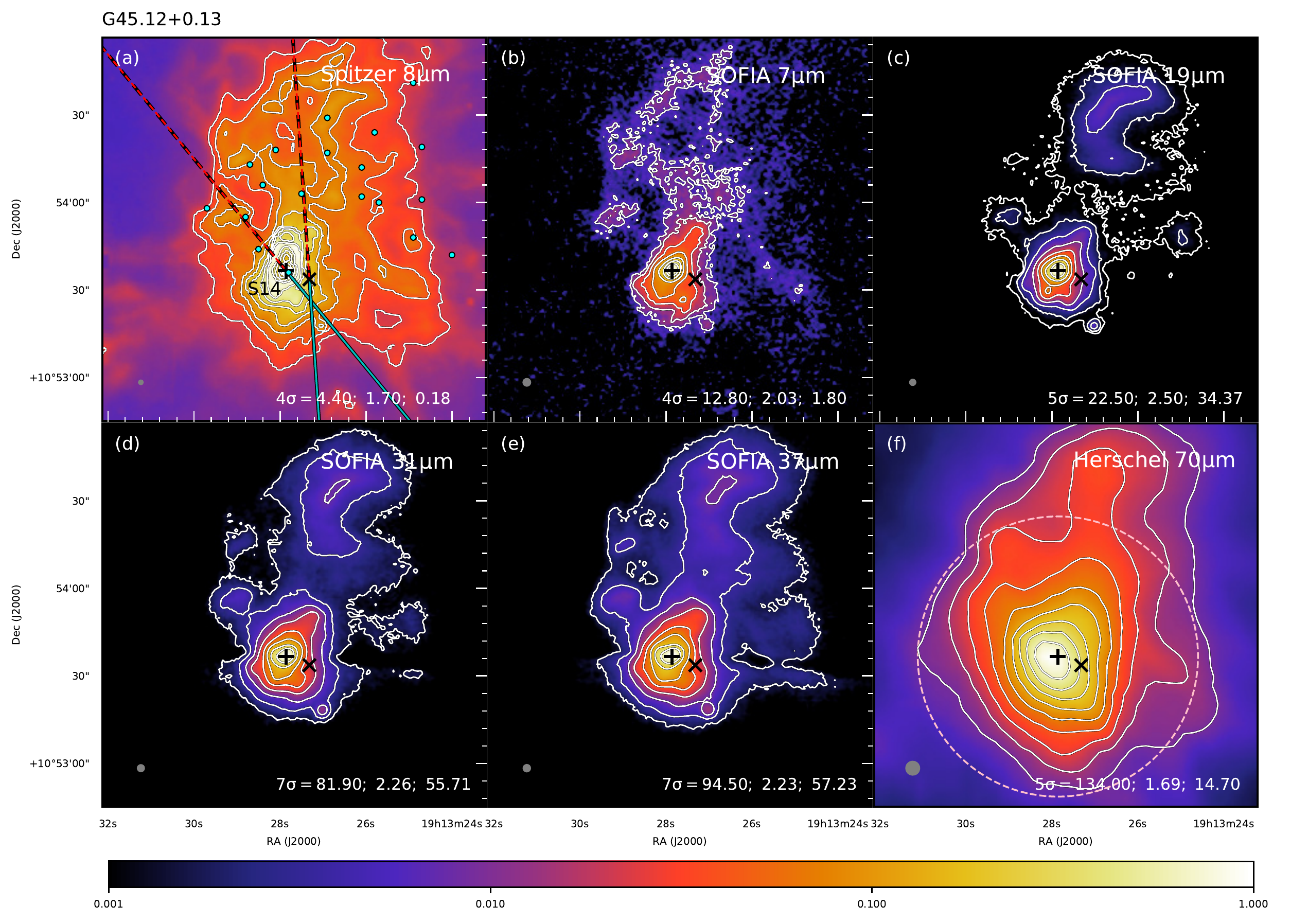}
\caption{
Multi-wavelength images of G45.12+0.13 with facility and wavelength
given in upper right of each panel.  Contour level information is
given in lower right: lowest contour level in number of $\sigma$ above
the background noise and corresponding value in mJy per square arcsec;
then step size between each contour in log$_{10}$ mJy per square
arcsec, then peak flux in Jy per square arcsec. The color map
indicates the relative flux intensity compared to that of the peak
flux in each image panel.
%but only showing the signal above 3$\sigma$. 
The pink dashed circle shown in (f) denotes the aperture used for the
fiducial photometry. Gray circles in the lower left show the
resolution of each image.  The black cross in all panels denotes the
peak position of the 6~cm continuum at R.A.(J2000) =
19$^h$13$^m$27$\fs$859, Decl.(J2000) =
$+$10$\arcdeg$53$\arcmin$36$\farcs$645 from Wood \& Churchwell
(1989). The $\times$ sign marks the suspected origin, G45.12+0.13
west, of one of the $^{13}$CO(1-0) outflows described in Hunter et
al. (1997). The lines in panel (a) show the orientation of outflow
axes, with the solid spans tracing blue-shifted directions and dashed
spans red-shifted directions. In this case, the outflow axis angles
are estimated from the $^{13}$CO(1-0) emission described in Hunter et
al. (1997). The cyan dots in panel (a) mark the 1.28\,GHz radio
continuum sources extracted in Vig et al. 2006. \label{fig:G45.12}}
\end{figure*}

This UC HII region, also known as IRAS 19111+1048, has a measured far
kinematic distance of 7.4~kpc (Ginsburg et al. 2011). The radio
morphology of this region shows a highly inhomogeneous ionized medium
(Vig et al. 2006), which is consistent with the extended MIR
morphology revealed here in Figure~\ref{fig:G45.12}.
%jct - this either needs better explanation, or just leave out
%The radio spectral types for majority of the compact sources match
%with the spectral type of the near-infrared counterparts.
Vig et al. (2006) proposed the source is an embedded cluster of Zero
Age Main Sequence (ZAMS) stars with twenty compact sources, including
one non-thermal source, identified by their radio emission. The
central UC HII source S14 is deduced to be of spectral type O6 from
the integrated radio emission. They also found there are two NIR
objects, IR4 and IR5, within the S14 region, while IR4 is at the peak
of the radio emission and matches the OH maser position obtained by
Argon et al. (2000). We see that most sources revealed at 8$\mu$m
and 37$\mu$m in the central region have counterparts in NIR bands (see
Figure~\ref{fig:nir}), which also indicates that this site is probably
a protocluster.

An extended bipolar outflow is revealed in CO(2--1), CO(3--2),
CO(6--5), $^{13}$CO(2--1) and C$^{18}$O(2--1) by Hunter et
al. (1997). Higher resolution $^{13}$CO(1--0) observations resolve the
system into at least two outflows. The highest velocity outflow
appears centered on the UC HII region S14. The additional bipolar
outflow was identified with a dynamical center lying offset
(-8$\arcsec$, -3$\arcsec$) from S14, named ``G45.12+0.13 west" by
Hunter et al. (1997). Hunter et al. (1997) argued that G45.12+0.13
west most likely represents dust emission from a younger or lower-mass
protostar that formed during the same epoch as the ionizing star of
S14. They also argued the absence of H$_2$O masers in the G45.12+0.13
cloud core suggests that both of the outflow sources have evolved
beyond the H$_2$O maser phase.

In our SOFIA images we see MIR to FIR emission peaking at the S14
position. We do not see a distinct source at the position of
G45.12+0.13 west, though the MIR extension to the southwest of S14
could be due to the two blue-shifted outflows, which are also revealed
in NIR (see Figure~\ref{fig:nir}). There is a MIR peak $\sim$
7.7\arcsec to the SE of S14, which is best revealed at 19\,$\mu$m and
further down $\sim$ 22\arcsec to the SW of S14 there is another MIR
peak. The closer one is seen in all J, H, K bands while the further
one is seen in H and K bands as shown in Figure~\ref{fig:nir}. They
could be more evolved low-mass protostars.

\subsubsection{G309.92+0.48}

\begin{figure*}
\epsscale{1.2}
\plotone{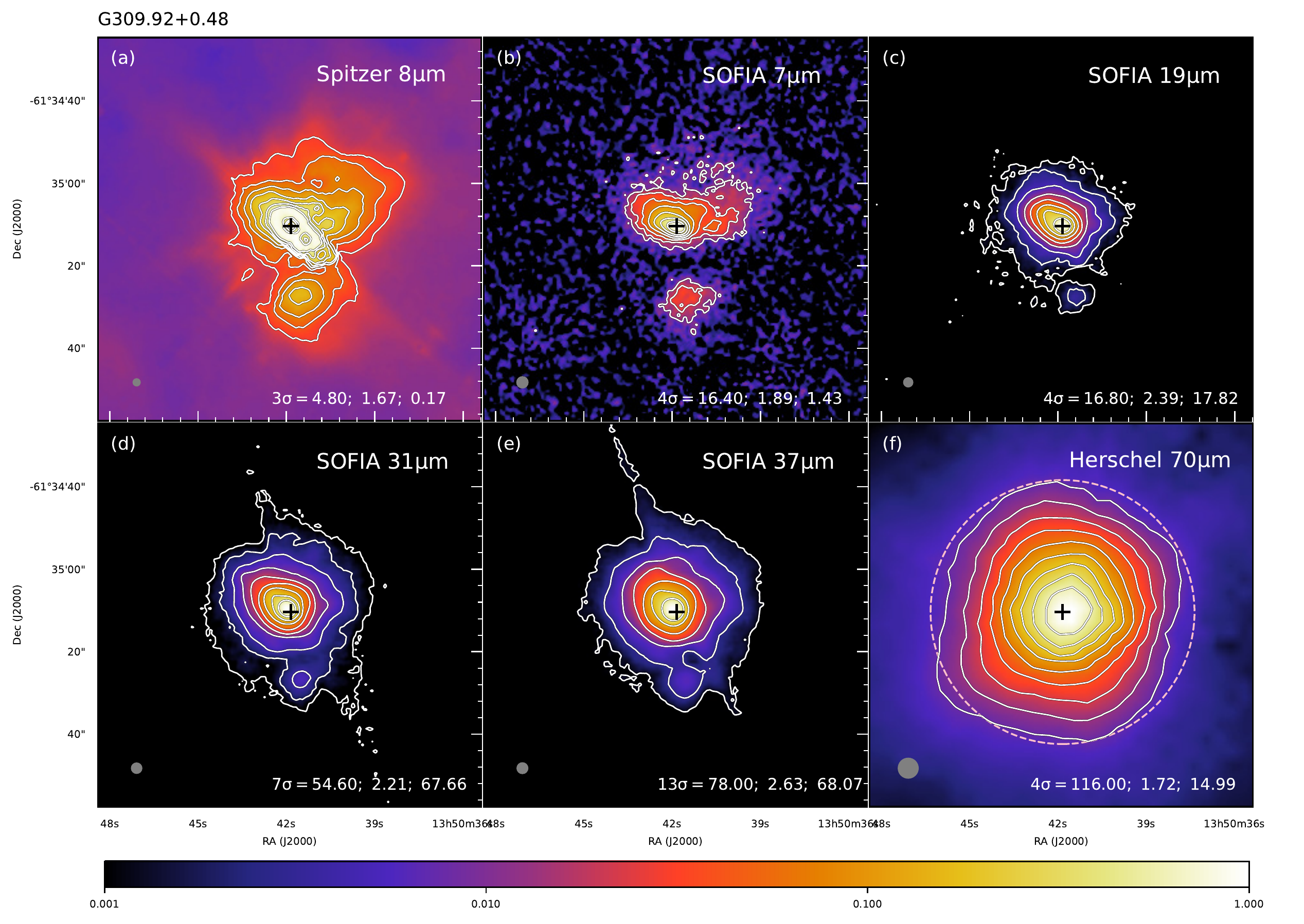}
\caption{
Multi-wavelength images of G309.92+0.48, following the format of
Figure 1. The black cross in all panels denotes the peak position of
the 8.6~GHz radio continuum estimated from Figure 5 in Philips et
al. (1998) at R.A.(J2000) = 13$^h$50$^m$41$\fs$847 ($\pm$0$\fs$015),
Decl.(J2000) = $-$61$\arcdeg$35$\arcmin$10$\farcs$40
($\pm$0$\farcs$12). Note that the extension of the central source to
the southwest in panel (a) is a ghosting effect, and not a real
structure. The stripes in panel (d) and (e) are also artifact features
caused by very bright point sources on the array. \label{fig:G309}}
\end{figure*}

\begin{figure}
\epsscale{1.15}
\plotone{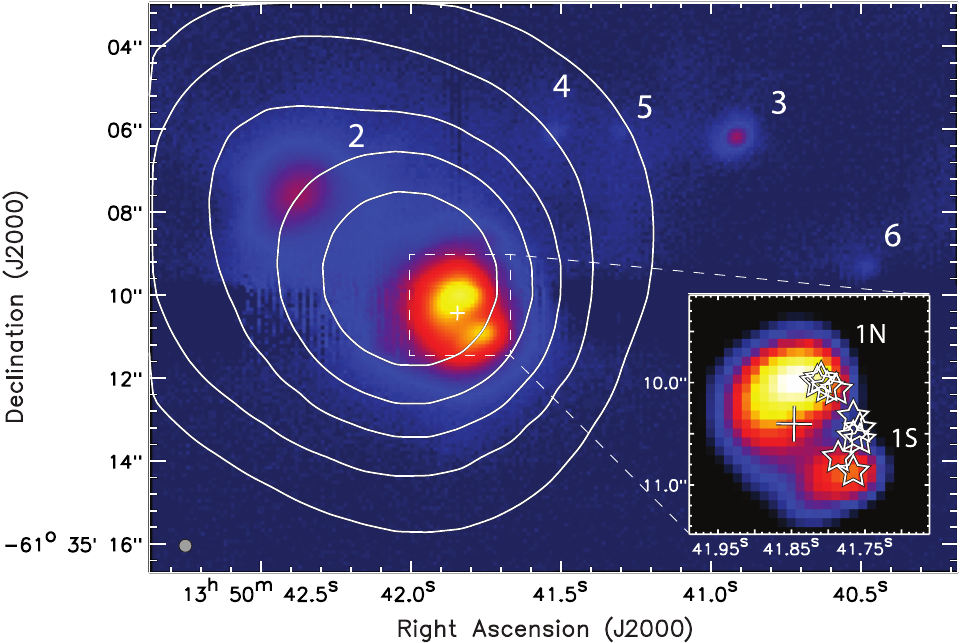}
\caption{
G309.92-0.48: color image is the \textit{Gemini} 11.7 $\mu$m image,
with IR source names labeled. The white contours are the SOFIA 37
$\mu$m data. The cross shows the peak location of the 8.6 GHz radio
continuum source of Phillips et al. (1998). The resolution of the
Gemini data is given by the gray circle in the lower left. The inset
shows a close-up of Source 1 at 11.7 $\mu$m, which is resolved into
two components labeled 1N and 1S. The radio continuum peak is again
shown as the cross, and the stars represent the locations of the 6.7
GHz methanol masers which form an arc-shaped distribution. Astrometry
between the radio masers (and continuum peak) and the 11.7 $\mu$m
image is better than 0.2\arcsec. Note that all the sources that appear
in the \textit{Gemini} field here are located within the northern
patch revealed by \textit{SOFIA} 7.7 $\mu$m in
Figure~\ref{fig:G309}. \label{fig:G309_Gemini}}
\end{figure}

This region is located at a distance of 5.5 kpc (Murphy et
al. 2010). The MIR emission in this area was resolved into 3 sources
with the CTIO 4-m at 10.8\,$\mu$m and 18.2\,$\mu$m, labeled 1 through
3 (see Figure 2 in De Buizer et al. 2000). In addition to these
sources, our \textit{Gemini} 11.7\,$\mu$m data also shows three
additional fainter point-sources, as shown in
Figure~\ref{fig:G309_Gemini}, which we label 4 through 6. Note that
all the sources that appear in the \textit{Gemini} field in
Figure~\ref{fig:G309_Gemini} are located within the northern patch
revealed by \textit{SOFIA} 7.7 $\mu$m in Figure~\ref{fig:G309}.

Source 1 is the brightest source in the MIR and is coincident with a
cm radio continuum source believed to be a HC HII region (Phillips et
al. 1998; Murphy et al. 2010).  Our \textit{Gemini} 11.7\,$\mu$m image
resolves Source 1 into two components as shown in
Figure~\ref{fig:G309_Gemini}, which we name 1N and 1S. Since both
sources are elongated at the same position angle, it may be that the
dark lane between them is an area of higher obscuration. In fact, the
radio continuum emission at 8.6~GHz (Walsh et al. 1998, Philips et
al. 1998) and 19~GHz (Murphy et al. 2010) towards Source 1 shows a
peak nearly in between mid-infrared Source 1N and 1S, possibly tracing
the location of the highly embedded protostar. Both of the radio
observations of Philips et al. (1998) and Murphy et al. (2010) show
elongation in the same direction as the MIR-dark lane. However, in
both cases the beam profile is also elongated in the same
direction. The 8.6~GHz observations of Walsh et al. (1998) have
similar resolution and a nearly circular beam, and do not show any
elongation.

OH and Class II methanol masers are found to be distributed
along an arc centered near the primary radio continuum peak (see inset
in Figure~\ref{fig:G309_Gemini}) with increasingly negative
line-of-sight velocities from north to south (Caswell 1997). Norris et
al. (1993) considered this site to have a well-defined methanol maser
velocity gradient and forwarded the idea that they are tracing a
near-edge-on circumstellar disk. The MIR morphology seen in the Gemini
data do not appear to support this idea. If the dark lane between
elongated Sources 1N and 1S is indeed the location of the protostar as
the radio peak suggests, then the morphology at 11.7\,$\mu$m would be
best explained as the emission from the walls of outflow cavities or
flared disk surfaces, with the dark lane representing a nearly
edge-on, optically-thick (in the IR), circumstellar disk. This disk
plane would be perpendicular to the methanol maser
distribution. Thus the Class II methanol masers may be coming
  from a region which experiences both strong shocks, but also a
  strong radiation field, which enables radiative pumping of the
  masers. To help infer the outflow orientation, De Buizer (2003)
observed the field for signs of H$_{2}$ emission, however, none was
detected (note, however, that this H$_{2}$ survey was relatively
shallow). We could not find any additional outflow information about
this region. Note that the extension of the central source in the
\textit{Spitzer} 8\,$\mu$m image and the stripes in the \textit{SOFIA}
31\,$\mu$m and 37\,$\mu$m images in Figure~\ref{fig:G309} are artifact
features caused by very bright sources on the array.

With the NIR VVV data, we find there is little to no NIR emission from
1N, which suggests that it is the most obscured source seen in the
MIR. In the J-band there is a compact emission source $\sim$2$\arcsec$
NE of the peak of Source 1N in the direction of Source 2, but no
emission directly coming from Source 1N or 1S. The H-band image shows
a source in this same location, but with the addition of an extended
source with a peak coincident with 1S, and a ``tail'' to the SE. At
Ks, there is only an extended source with a peak at 1S, and extended
emission in the same direction as the tail seen in H-band, with
emission also extending NE towards 1N. Source 2 lies to the northeast
of Source 1 at a position angle of 53$\arcdeg$. Both Source 1 and 2
are seen at 8.6 GHz by Phillips et al. (1998) and in the NIR by Walsh
et al. (1999). With the NIR VVV data, we find that Sources 2 and 3 are
also seen at J, H and Ks. Source 6 is also seen at J, H, and Ks,
Source 4 is seen at H and Ks, but Source 5 is not detected in the
NIR. In our 7.7\,$\mu$m \textit{SOFIA} data we see fingers of emission
reaching the area around Sources 3 and 5, as well as Source 6, though
these are not detected at longer wavelengths in the \textit{SOFIA}
data.

In the larger field of view of the \textit{SOFIA} data, we detect
another extended (r$\sim$5$\arcsec$) emission region $\sim$18$\arcsec$
south of Source 1 at all \textit{SOFIA} wavelengths. The nature of
this region is unknown, however.

\subsubsection{G35.58-0.03}
\begin{figure*}
\epsscale{1.2}
\plotone{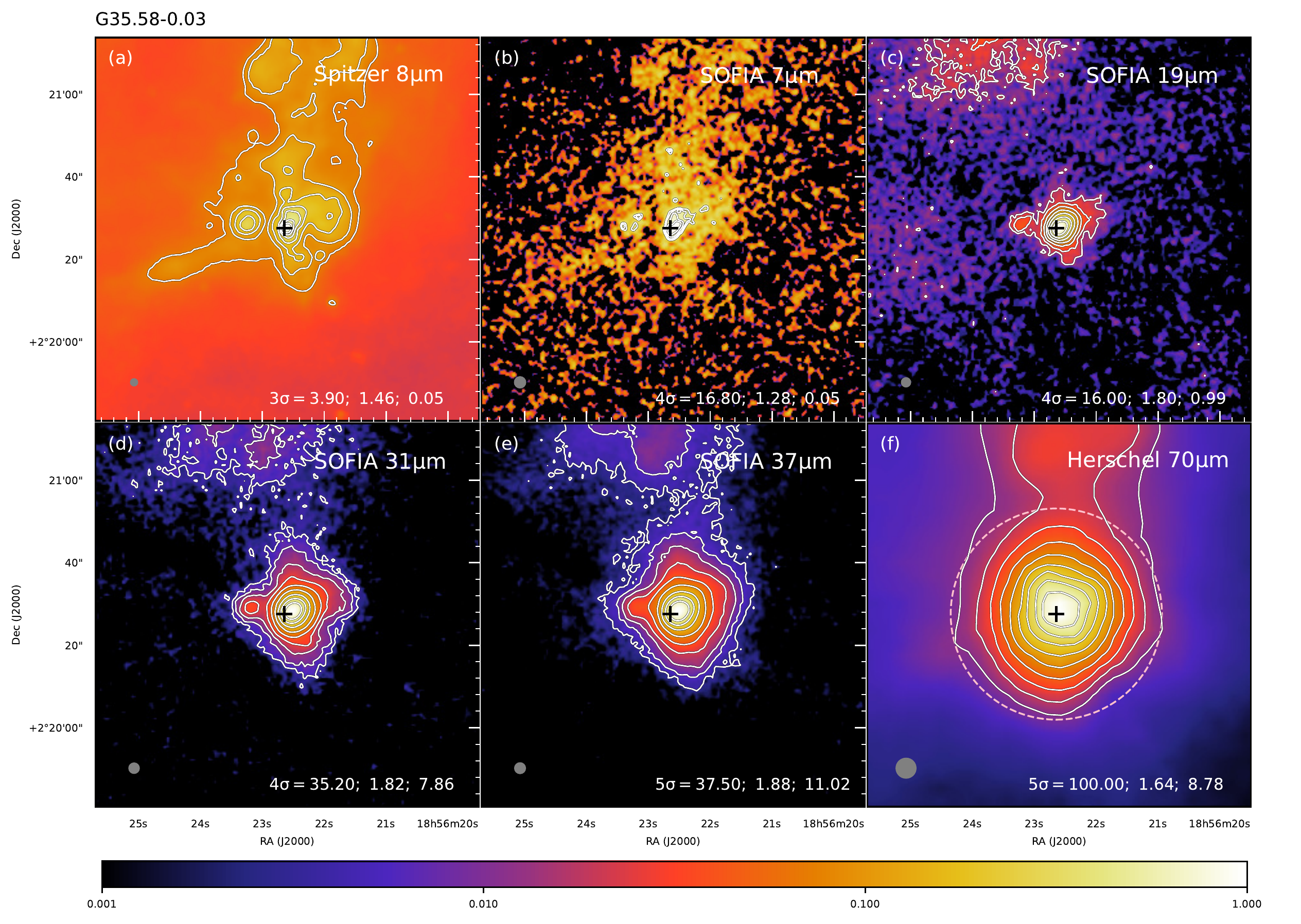}
\caption{
Multi-wavelength images of G35.58-0.03, following the format of Figure
1. 
%The black cross in all panels denotes the peak position of the UC HII region from Zhang et al. (2014) VLA 3.6~cm continuum emission at R.A.(J2000) = 18$^h$56$^m$22$\fs$563, Decl.(J2000) = $+$02$\arcdeg$20$\arcmin$27$\farcs$660. 
The black cross in all panels denotes the peak position of the UC HII
region G35.578-0.031 from Kurtz et al. (1994) 2\,cm radio continuum
emission at R.A.(J2000) = 18$^h$56$^m$22$\fs$644, Decl.(J2000) =
$+$02$\arcdeg$20$\arcmin$27$\farcs$559.  \label{fig:G35.58}}
\end{figure*}

\begin{figure}
\epsscale{1.15}
\plotone{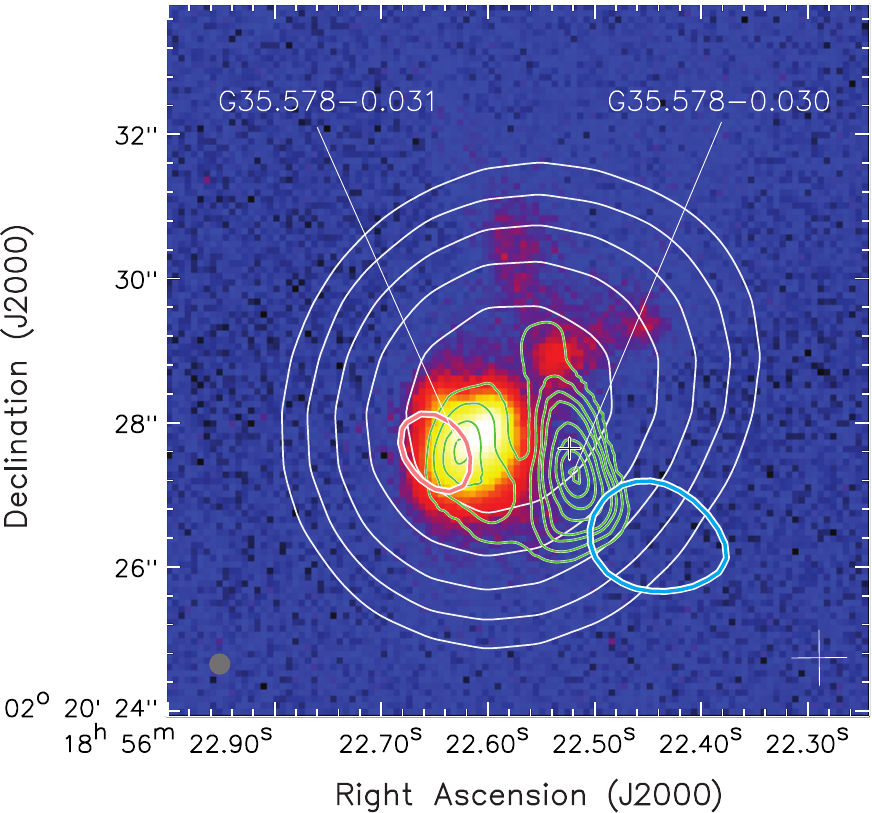}
\caption{
G35.58-0.03. The color image is the \textit{Gemini} 11.7\,$\mu$m
data. The white contours are the \textit{SOFIA} 37\,$\mu$m data. The
green contours are the 2\,cm radio continuum emission as seen by Kurtz
et al. (1994), and the names of the two radio sources are labeled. The
black cross shows the peak location of the ammonia and 1.3\,cm radio
continuum source of Zhang et al. (2014). The size of this cross also
denotes the astrometric error between the between all of the radio
data and the 11.7\,$\mu$m image (0.3$\arcsec$). The red and blue
contours are the brightest red- and blue-shifted CO(2--1) outflow
contours from Zhang et al. (2014). The resolution of the
\textit{Gemini} data is given by the gray circle in the lower left
corner. The astrometry uncertainty between the {\it SOFIA} 37\,$\mu$m
contours and the radio data are given by the white cross in the lower
right corner. \label{fig:G35.58_Gemini}}
\end{figure}

The star-forming region G35.58-0.03 is located at the far kinematic
distance of 10.2 kpc (Fish et al. 2003; Watson et al. 2003). Kurtz et
al. 1994 resolved the 2\,cm and 3.6\,cm continuum emission here into
two UC HII regions $\sim$2$\arcsec$ apart, with the western source
named G35.578-0.030 and the eastern source named
G35.578-0.031. G35.578-0.030 contains water and OH masers, but no
methanol masers (Caswell et al. 1995). Zhang et al. (2014) found that
there is an ammonia clump peaked co-spatially with their observed
1.3\,cm radio continuum peak, which is $\sim$0.4$\arcsec$ north of the
2\,cm peak of G35.578-0.030 (Kurtz et al. 1994; 1999). H30$\alpha$
shows evidence of an ionized outflow connecting to a molecular outflow
seemingly centered on the radio continuum peak of G35.578-0.030. 
Only faint 1.3\,cm continuum emission was found from the eastern source,
G35.578-0.031, and no signs of outflow or ammonia emission.

De Buizer et al. (2005) presented $\sim$0.6$\arcsec$ resolution
MIR images of this region at 10 and 20\,$\mu$m, which showed
a single source with some extension to the northwest. Due to poor
astrometry of the data, it was unclear which UC HII region the
mid-infrared emission was associated with. They argued that, due to
the fact that the western source, G35.578-0.030, appears to have a
similar extension to the northwest at 3.6\,cm as seen by Kurtz et
al. (1999), that the MIR emission is likely to be associated
with that source.

Our data obtained at 11.7\,$\mu$m from \textit{Gemini} with
$\sim$0.3$\arcsec$ resolution further resolve the MIR emission into a
main bright peak with two fingers of extended diffuse emission to the
north and northwest. Using \textit{Spitzer} 8\,$\mu$m images to
confirm our astrometry, it is revealed that the MIR peak is not
associated with the western UC HII region, but instead the eastern UC
HII region, G35.578-0.031 (see Figure~\ref{fig:G35.58_Gemini}). The
relative astrometric error between the {\it Gemini} 11.7\,$\mu$m image
and the radio data is better than 0.3$\arcsec$. No MIR emission is
detected at the location of G35.578-0.030 out to 37\,$\mu$m. The MIR
peak is, however, close to the location of the redshifted outflow
cavity of G35.578-0.030 seen in CO(2--1) by Zhang et
al. (2014). However, if high extinction was causing the general lack
of MIR emission from G35.578-0.030, it seems unlikely that the MIR
emission we are seeing would come from the even more extinguished
red-shifted outflow cavity of G35.578-0.030. It is more plausible that
the MIR emission is coming solely from the eastern UC HII region,
G35.578-0.031.

Our \textit{SOFIA} images of this region (Figure~\ref{fig:G35.58})
show a bright source peaked at the location of G35.578-0.031 and
extended slightly to the northwest, as is seen in the higher spatial
resolution \textit{Gemini} 11.7\,$\mu$m image
Figure~\ref{fig:G35.58_Gemini}. The nature of this extension is
unclear, since the outflow seen by Zhang et al. (2014) has an axis
oriented east-west. A second compact source is detected in our
\textit{SOFIA} data (and in the \textit{Spitzer}-IRAC data) located
$\sim$10$\arcsec$ to the east of G35.578-0.031. There is also a hint
of MIR extension to the west, which may be due to the outflow.
%Its emission becomes weaker as the wavelength increases. At 19, 31, and 37\,$\mu$m, the flux from the eastern component is less than 5\% than that from the main source. From our \textit{SOFIA} MIR images, the extended emission to the east of the brightest peak is most likely due to the secondary source.

The eastern MIR source seen in the \textit{SOFIA} data has a
counterpart at K-band as can be seen from Figure~\ref{fig:nir}. Thus
it may be a more evolved protostar, closer to the end of its
accretion. From the NIR image (see Figure~\ref{fig:nir}) there are at
least two K-band sources within the highest contour of the 37\,$\mu$m
emission. The southern K-band source is associated with the peak at
8\,$\mu$m and the main bright peak at 11.7\,$\mu$m, while the northern
K-band source has some overlap with the northern finger in
\textit{Gemini} 11.7\,$\mu$m image (not shown here). There could be
one or two lower luminosity companion sources in that region together
with the southern main massive protostar, but they are not well
resolved in the MIR and FIR.

\subsubsection{IRAS 16562-3959}

\begin{figure*}
\epsscale{1.2}
\plotone{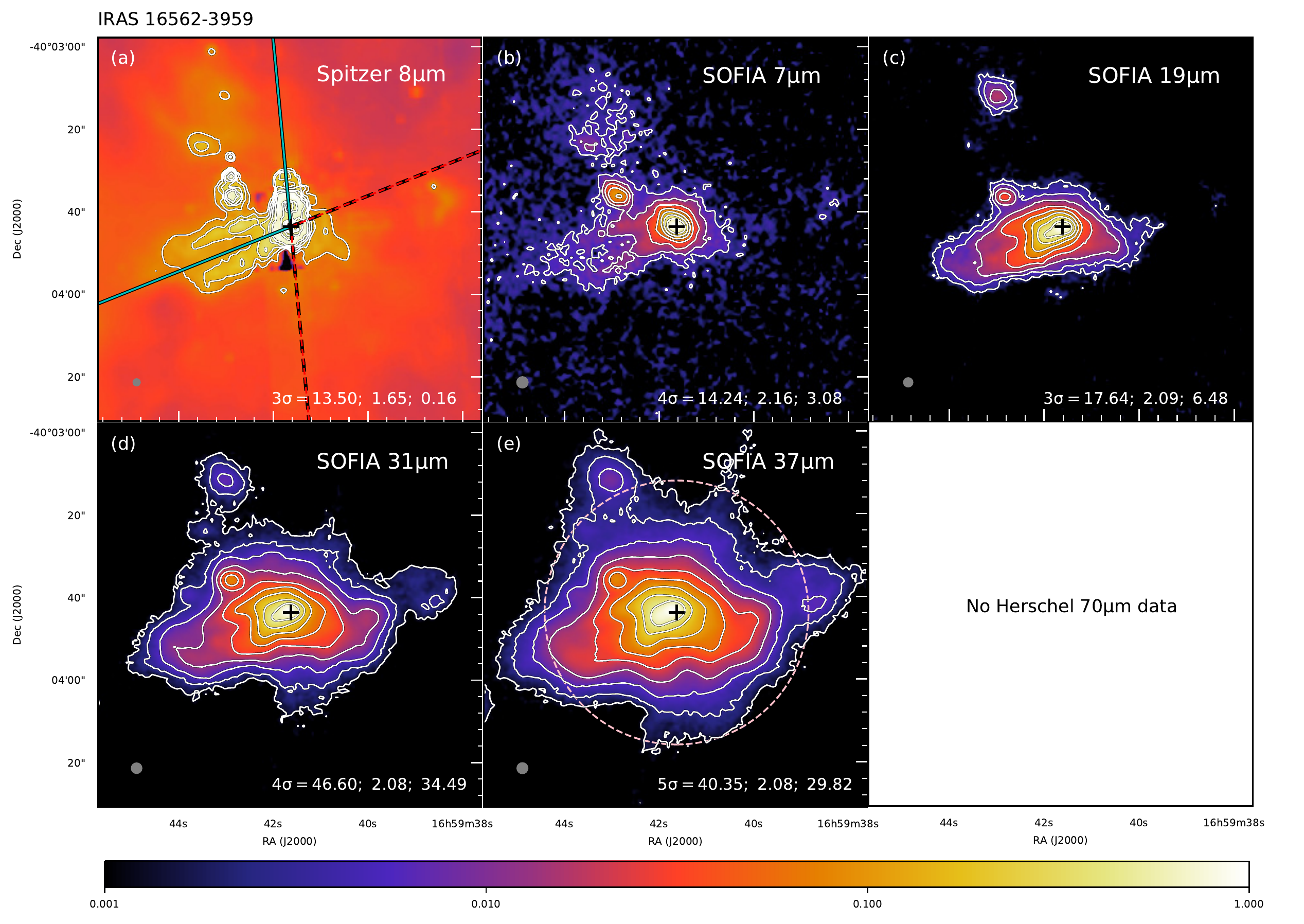}
\caption{
Multi-wavelength images of IRAS~16562, following the format of Figure
1.  The black cross in all panels denotes the position of the central
8.6 GHz radio source (C) from Guzman et al. (2010) at R.A.(J2000) =
16$^h$59$^m$41$\fs$63, Decl.(J2000) =
$-$40$\arcdeg$03$\arcmin$43$\farcs$61. The lines in panel (a) show the
outflow axis angles, with the solid spans tracing the blue-shifted
directions and dashed spans the red-shifted directions. The
outflow axis angles are from the CO(6-5) emission of Guzman et
al. (2011). Note the extension and the dark appearance at the center
in panel (a) are ghosting effects.\label{fig:IRAS16562}}
\end{figure*}

This source (also known as G345.49+1.47) is located at a distance of
1.7~kpc (Guzm\'an et al. 2010). It is believed that the massive core
hosts a high-mass star in an early stage of evolution, including
ejection of a powerful collimated outflow (Guzm\'an et
al. 2010). Guzm\'an et al. (2010) carried out ATCA observations to
reveal five 6~cm radio sources: a compact bright central (C)
component, two inner lobes that are separated by about 7$\arcsec$ and
symmetrically offset from the central source, and two outer lobes that
are separated by about 45$\arcsec$ (see Figure 4 in Guzman et
al. 2010). The central radio source has a 3\,mm counterpart, source 10
in Guzm\'an et al. (2014), and an X-ray counterpart, source 161 in
Montes et al. (2018), and is associated with OH maser emission
(Caswell 1998, 2004). It is interpreted as a HC HII region based on
hydrogen recombination line (HRL) observations (Guzm\'an et
al. 2014). The continuum at 218 GHz and CH$_3$CN(12--11)
(methylcyanide) observations by Cesaroni et al. (2017) revealed that
the central source 10 actually consists of two peaks. The four other
symmetrically displaced sources are interpreted as shock-ionized lobes
(Guzm\'an et al. 2010) and are observed to move away from the central
source at high speed (Guzm\'an et al. 2014).

On the other hand, the molecular observations of CO(6--5) and CO(7--6)
show the presence of high-velocity gas exhibiting a quadrupolar
morphology (Guzm\'an et al. 2011), most likely produced by the
presence of two collimated outflows, one major outflow lying with a
southeast-northwest (SE-NW) orientation, and the other with a N-S
orientation, which may come from the unresolved mm source 13 in Guzm\'an
et al. (2014) to the east of the central source. The SE-NW molecular
outflow is aligned with the string of radio continuum
sources. Extended Ks-band emission probably tracing excited $\rm
H_2$-2.12 $\mu$m is also associated with the SE-NW flow.

In Guzm\'an et al. (2014), the molecular core in which the outflow is
embedded presents evidence of being in gravitational contraction as
shown by the blue asymmetric peak seen in HCO$^+$(4--3). The emission
in the SO$_2$, $^{34}$SO, and SO lines exhibits velocity gradients
interpreted as arising from a rotating compact ($\sim$ 3000\,AU)
molecular core with angular momentum aligned with the jet
axis. L{\'o}pez-Calder{\'o}n et al. (2016) reported $^{13}$CO(3--2)
APEX observations of this region and showed that the high-mass
protostellar candidate is located at the column density
maximum. Montes et al. (2018) decomposed the wider region into 11
subclusters with results from {\it Chandra} X-ray observations
together with VISTA/VVV and {\it Spitzer}-GLIMPSE catalogs and the
subcluster containing the high-mass protostar was found to be the
densest and the youngest in the region with the high-mass protostar
located near its center.

In our MIR images, the extended IR emission is likely tracing the
illuminated inner outflow cavity containing the jet. There are two
knots to the northeast of the central source revealed by {\it SOFIA}. The
closer knot located $\sim$15$\arcsec$ NE of the central source
is associated with the 92.3 GHz peak 18 in Guzm\'an et al. (2014), as
well as a K-band source (see Figure~\ref{fig:nir}). It may correspond
to the X-ray source 178 in Montes et al. (2018). There is OH maser
emission (Caswell 1998, 2004), but no radio continuum emission
detected. Thus it may be a low-mass protostar. The farther knot,
located $\sim$36$\arcsec$ northeast of the central source, has
counterparts in all of the J, H, K bands. We did not find any
associated X-ray source for this knot in the Montes et al. (2018)
sample.

\subsubsection{G305.20+0.21}

\begin{figure*}
\epsscale{1.2}
\plotone{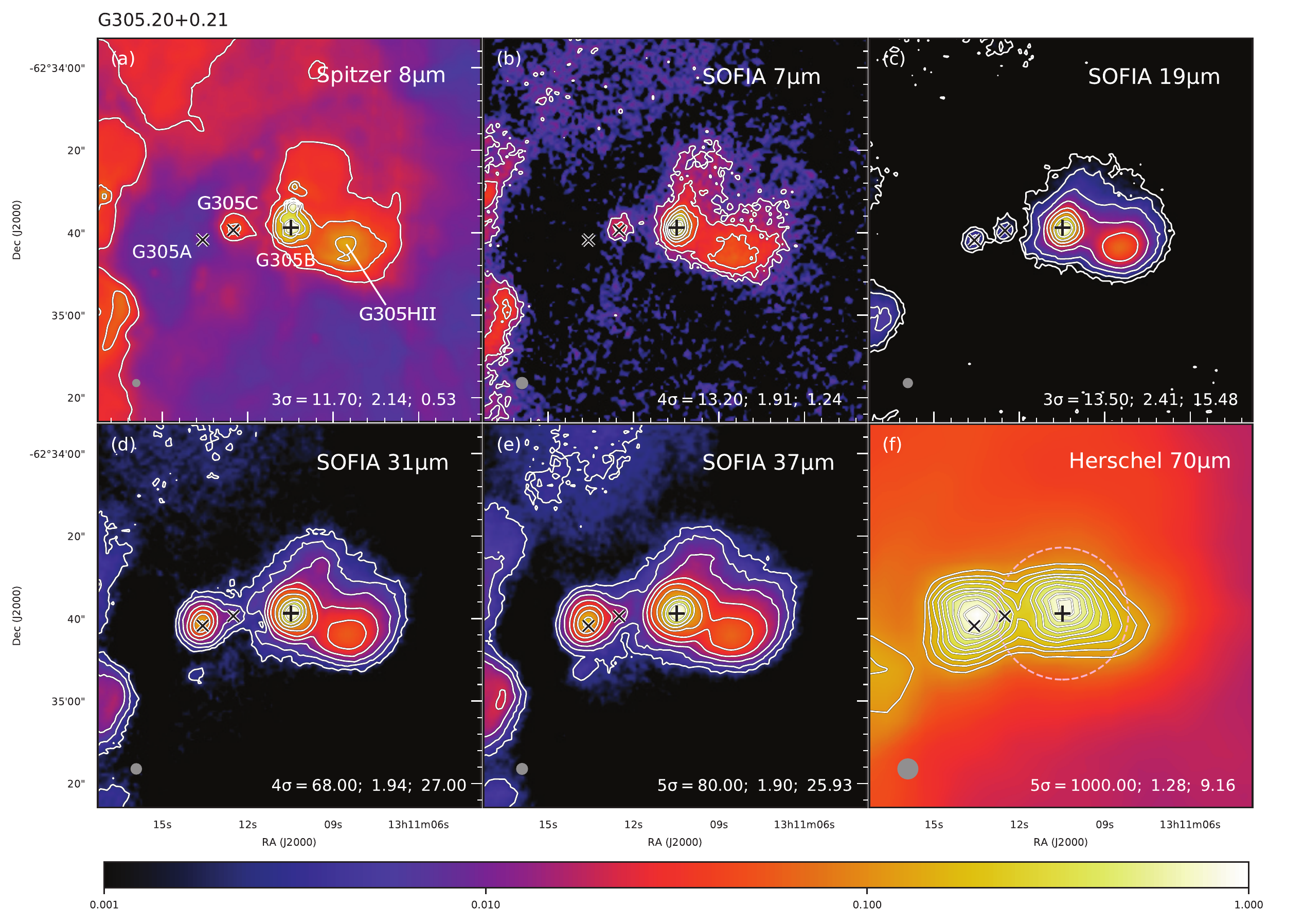}
\caption{
Multi-wavelength images of G305.20+0.21, following the format of
Figure 1. The black cross in all panels denotes the peak position of
the 6.7 GHz methanol maser from Caswell, Vaile \& Forster (1995b) at
R.A.(J2000) = 13$^h$11$^m$10$\fs$49, Decl.(J2000) =
$-$62$\arcdeg$34$\arcmin$38$\farcs$8. The $\times$ signs denote the
MIR peak positions of G305A and G305C determined from the {\it SOFIA}
19\,$\mu$m image. \label{fig:G305}}
\end{figure*}

\begin{figure*}
\epsscale{1.15}
\plotone{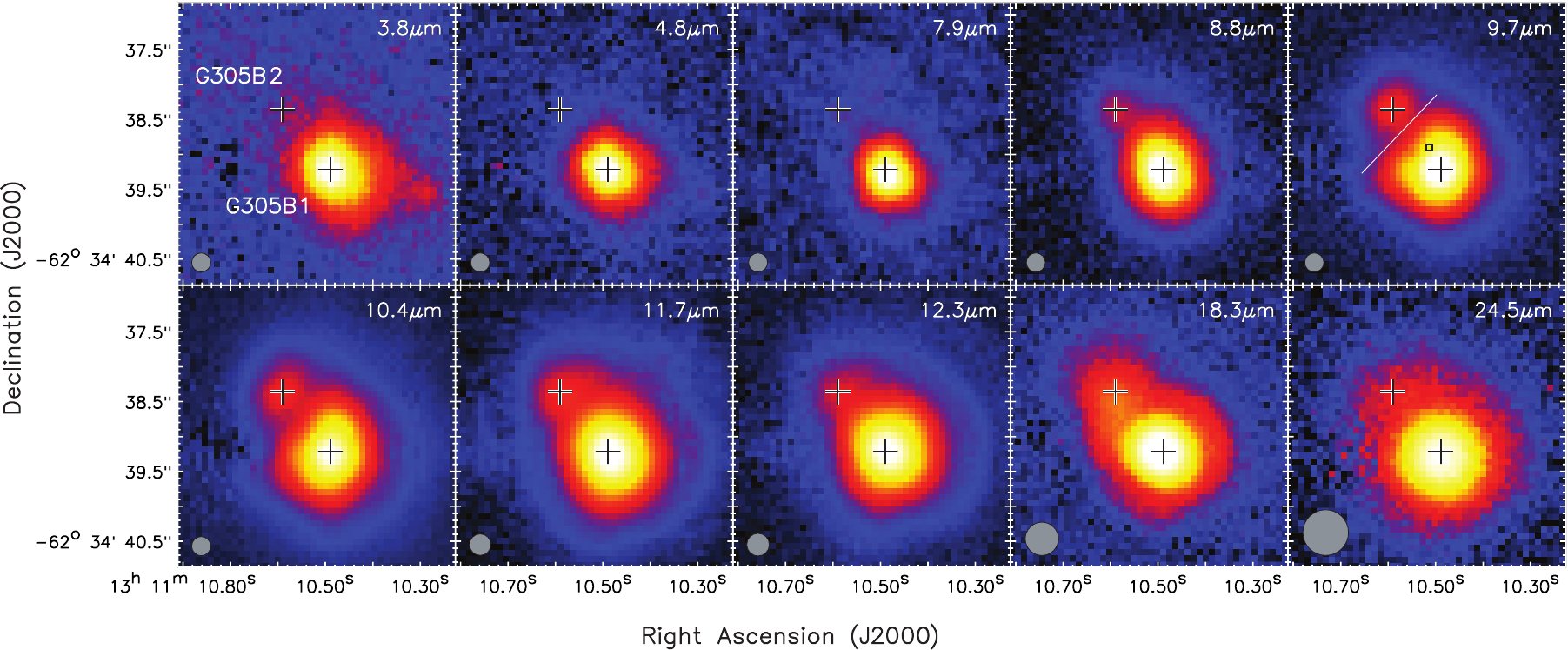}
\caption{
G305.20+0.21. We present \textit{Gemini} images at 10 different MIR
wavelengths from 3.8 to 24.5\,$mu$m. The wavelength of the image is
given in the upper right corner of each panel and the resolution is
given by the gray circle in the lower left corner of each
panel. Infrared source names are labeled in the top left panel, and
their peak locations (as determined from the 9.7\,$\mu$m image) are
given in each panel by the crosses. The square in the upper right
panel represents the location of the 6.7 GHz methanol maser reference
feature of Phillips et al. (1998). Astrometry between the maser
location and the \textit{Gemini} data is better than 0.2\arcsec. The
white line in the upper right panel is present to demonstrate the flatness
of the northeast side of G305B1. \label{fig:G305_Gemini}}
\end{figure*}

G305.20+0.21 is a massive star-forming region located at a distance of
$4.1^{+1.2}_{-0.7}$ kpc from parallax of 6.7 GHz methanol masers
(Krishnan 2017). Class II methanol (CH$_{3}$OH) masers were reported
in two positions by Norris et al. (1993): G305.21+0.21 and
G305.20+0.21 separated by approximately 22$\arcsec$. Walsh \& Burton
(2006) refer to these maser sites as G305A and G305B, respectively,
and we will adopt that nomenclature here.

The brightest MIR source appears to be associated with the methanol
masers of G305B, but does not possess detectable radio continuum
emission (below a 4$\sigma$ detection limit of 0.9 mJy\,beam$^{-1}$
(beam $\sim$ 1.5$\arcsec$) at 8.6\,GHz in Phillips et al. 1998, and a
3$\sigma$ detection limit of 0.09\,mJy at 18\,GHz in Walsh et
al. 2007). Walsh et al. (2007) found no HC$_3$N, NH$_3$, OCS, or water
at the position of G305B and proposed that it has evolved enough to
the point that it has already had time to clear out its surrounding
molecular material. By contrast, Boley et al. (2013) proposed G305B is
a massive protostar in a pre-UCHII-region stage. Our \textit{SOFIA}
images show that G305B is the brightest MIR source out to
37\,$\mu$m. Our high-spatial-resolution \textit{Gemini} data
(Figure~\ref{fig:G305_Gemini}) show G305B is resolved into two
emission components, with the fainter secondary source (which we name
G305B2) lying $\sim$1$\arcsec$ to the NE of the brighter source
(G305B1). G305B2 is only visible at wavelengths greater than
8.8\,$\mu$m. By contrast, G305B1 is seen to have emission in all {\it
  Gemini} images from 3.8 to 24.5\,$\mu$m, and has a NIR counterpart
as well (see Figure~\ref{fig:nir} and Walsh et al. 1999). Using four
infrared sources seen in both the {\it Gemini} 3.8\,$\mu$m image (but
not shown in Figure~\ref{fig:G309_Gemini}) and the Spitzer 3.6$\mu$m
image we were able to confirm the absolute astrometry of the {\it
  Gemini} data at all wavelengths to better than 0.2$\arcsec$. This
places the Class II methanol maser reference feature (i.e., the
brightest maser spot) from Phillips et al. (1998) $\sim$0.5\arcsec
NE of the MIR peak (see the 9.7$\mu$m image in
Figure~\ref{fig:G305_Gemini}). It is not clear what these masers are
tracing.

What is the nature of the MIR double source associated with
G305B? G305B2 could be a more embedded source, since it is not visible
at shorter IR wavelengths. However, it appears to change shape
considerably as a function of wavelength, flattening and becoming more
diffuse at 18.3 and 24.5\,$\mu$m. G305B1 also changes shape modestly
with wavelength and its shape at 9.7 and 10.4\,$\mu$m is peculiar. The
northeast side of G305B1 is very flat, and almost completely straight
at 9.7 and 10.4\,$\mu$m (see white line in the 9.7$\mu$m panel of
Figure~\ref{fig:G305_Gemini} as reference). As these filters are
sampling the wavelength of peak dust extinction (Gao et al.
2009), it may be that the morphologies of both sources could be
explained if the dark lane between them is a ``silhouette'' of a
circumstellar disk or toroid that is optically thick in the MIR. The
brighter MIR source G305B1 would be the side of the disk or outflow
cavity facing towards us, and G305B2 the side facing away which we
only see at longer wavelengths due to extinction from the disk along
the line of sight. We could corroborate the outflow cavity hypothesis
if we had evidence of an outflow and knew its angle. Walsh et
al. (2006) did image the area in commonly used outflow tracers
$^{13}$CO and HCO$^+$, and presented the data as integrated emission
maps. However, the emission appears to peak on G305A and extends at
larger scales in a direction parallel to the dark lane orientation,
tracing the location of the extended 1.2\,mm continuum emission
(rather than an outflow). However, if the hypothesis of Walsh et
al. (2007) is correct, i.e., that due to low chemical abundance
this source is more evolved and has cleared much of its surrounding
molecular material, then the source may have passed the stage where it
would exhibit an active outflow. Conversely, a Class I methanol maser
was detected by Walsh et al. (2007) 3$\arcsec$ due east of G305B, and
they are generally only found in outflows.

Walsh et al. (2001) observed the 6.7\,GHz methanol maser site G305A
in the MIR (10.5\,$\mu$m and 20\,$\mu$m) and found that G305A is not
associated with any MIR source. G305A is out of the field of our
\textit{Gemini} images. However, we see strong emission from G305A in
our \textit{SOFIA} images at 19\,$\mu$m and longer, and it becomes the
dominant source in the FIR starting at \textit{Herschel}
70\,$\mu$m. G305A is also not associated with any 8.6\,GHz continuum
emission with a flux density limit of 0.55\,mJy beam$^{-1}$ (Phillips
et al. 1998) or 18 GHz continuum emission with a detection limit of
0.15\,mJy (Walsh et al. 2007), but is rich in molecular tracers (Walsh
et al. 2007) indicating it is a source that is likely much younger and
more embedded than G305B and in a hot core phase, prior to the onset
of a UC HII region.

About 15$\arcsec$ to the southwest of G305B is an extended HII region,
G305HII, with a flux of 195\,mJy at 8.6\,GHz (Phillips et
al. 1998). We detect this source in all of our \textit{SOFIA}
images. We also detect an infrared source between G305A and G305B,
which we call G305C, located $\sim$ 14\arcsec east of G305B. It is
present at all wavelengths in the \textit{SOFIA} images, but becomes
less pronounced at longer wavelengths. It also has NIR counterparts,
as shown in Figure~\ref{fig:nir}, which seem to be resolved into three
peaks. The nature of the source is uncertain, but it may be a low mass
YSO. Besides the G305HII region there is no other radio emission in
the field shown in Figure~\ref{fig:G305} revealed by the 18 GHz
continuum in Walsh et al. (2007).

\subsubsection{G49.27-0.34}

\begin{figure*}
\epsscale{1.2}
\plotone{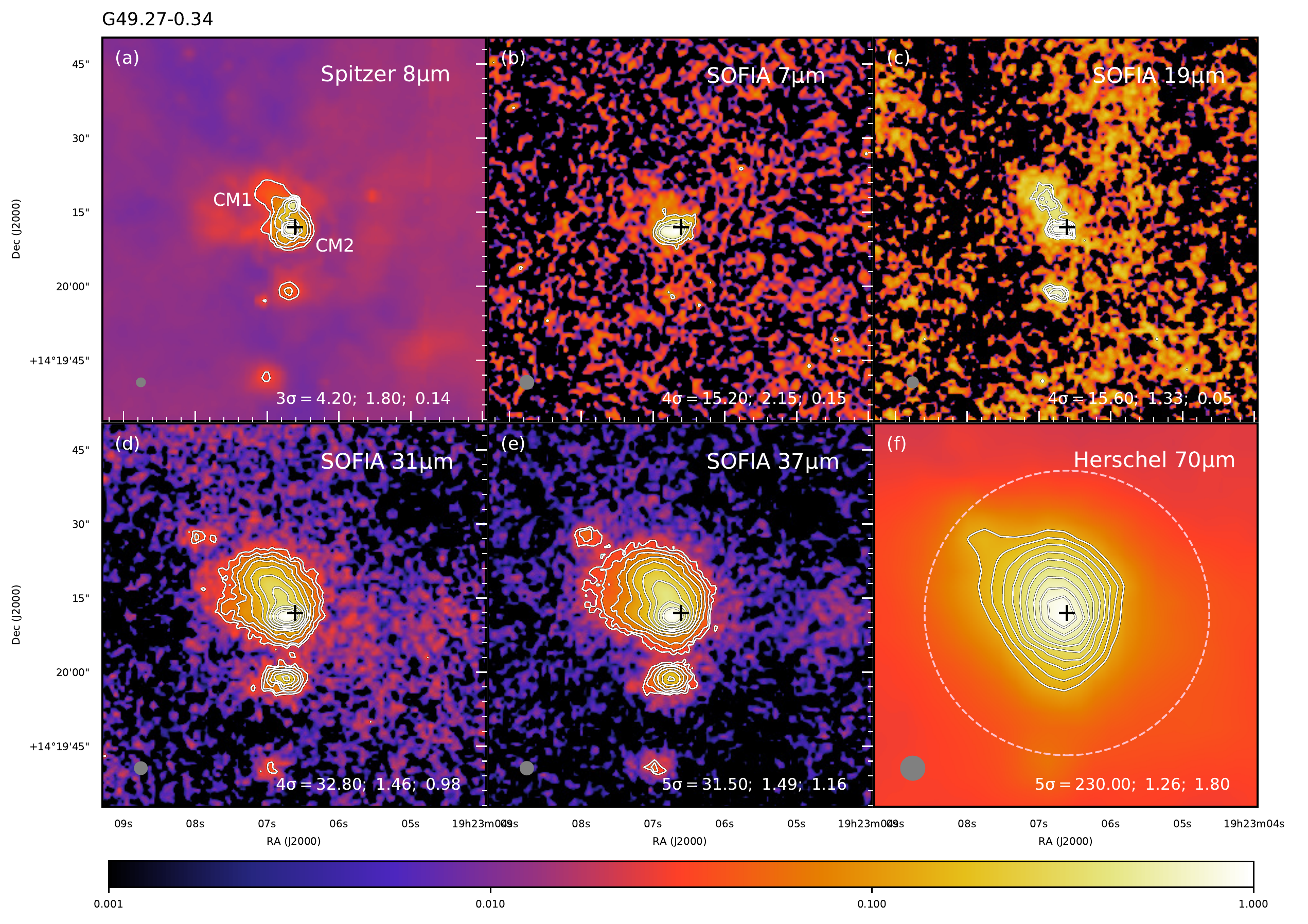}
\caption{
Multi-wavelength images of G49.27-0.34, following the format of Figure
1. The black cross in all panels denotes the peak position CM2 of the
3.6 cm continuum from Cyganowski et al. (2011a) at R.A.(J2000) =
19$^h$23$^m$06$\fs$61, Decl.(J2000) =
$+$14$\arcdeg$20$\arcmin$12$\farcs$0. \label{fig:G49.27}}
\end{figure*}

This source, classed as an ``extended green object'' (EGO) is in an
IRDC with near kinematic distance of 5.55 $\pm$ 1.66\,kpc (Cyganowski
et al. 2009). The MIR peak (see Figure~\ref{fig:G49.27}) is associated
with the 3.6\,cm radio source CM2 in Cyganowski et al. (2011a). Towner
et al. (2017) did not detect a 1.3 cm counterpart to CM2 at the a $4
\sigma$ detection limit of 0.28 mJy beam$^{-1}$ (beam $\sim$
1\arcsec). The MIR extension to the northeast is associated with a
stronger radio source CM1 detected at 3.6\,cm and 1.3\,cm by
Cyganowski et al. (2011a) and at 20\,cm by Mehringer (1994).

%The 3.6 and 1.3 cm flux densities of CM1 are consistent with optically thin free-free emission; the ionizing photon flux ($\sim2.2 \times 10^{47} \rm s^{-1}$) corresponds to a single ionizing star of spectral type B0V, in agreement with the estimate of Mehringer (1994) based on an unresolved 20 cm detection ($\sim$14" resolution). The spectral index limit for CM2 ($<$ 0.2) is also consistent with optically thin free-free emission, but the calculated ionizing photon flux is about two orders of magnitude lower ($\sim1.79 \times 10^{45} \rm s^{-1}$) (Cyganowski et al. 2011a).

We did not find any outflow information about this source. De Buizer
\& Vacca (2010) obtained \textit{Gemini} L- and M-band spectra for
this EGO, and detected only continuum emission (no H$_2$ or
CO). However, Cyganowski et al. (2011a) suspected that an outflow,
perhaps driven by CM2 or by a massive protostar undetected at cm
wavelengths, may exist, but is not detected, given the 44\,GHz Class I
CH$_3$OH masers and 4.5\,$\mu$m emission in the south. SiO(5--4),
HCO$^+$ and H$^{13}$CO$^+$ emission is detected toward this EGO with
JCMT (Cyganowski et al. 2009). No 6.7\,GHz CH$_3$OH emission is
detected towards this EGO (Cyganowski et al. 2009). Neither thermal
nor maser 25 GHz CH$_3$OH emission is detected (Towner et al. 2017).

There is a secondary component revealed by our \textit{SOFIA} data to
the south of the main MIR peak. It is also seen at 3.6\,cm (Cyganowski
et al. 2011a) but not at 1.3\,cm (Towner et al. 2017). The nature of
this source is unknown. We do not see obvious counterparts in the NIR
image (see Figure~\ref{fig:nir}).

\subsubsection{G339.88-1.26}

\begin{figure*}
\epsscale{1.2}
\plotone{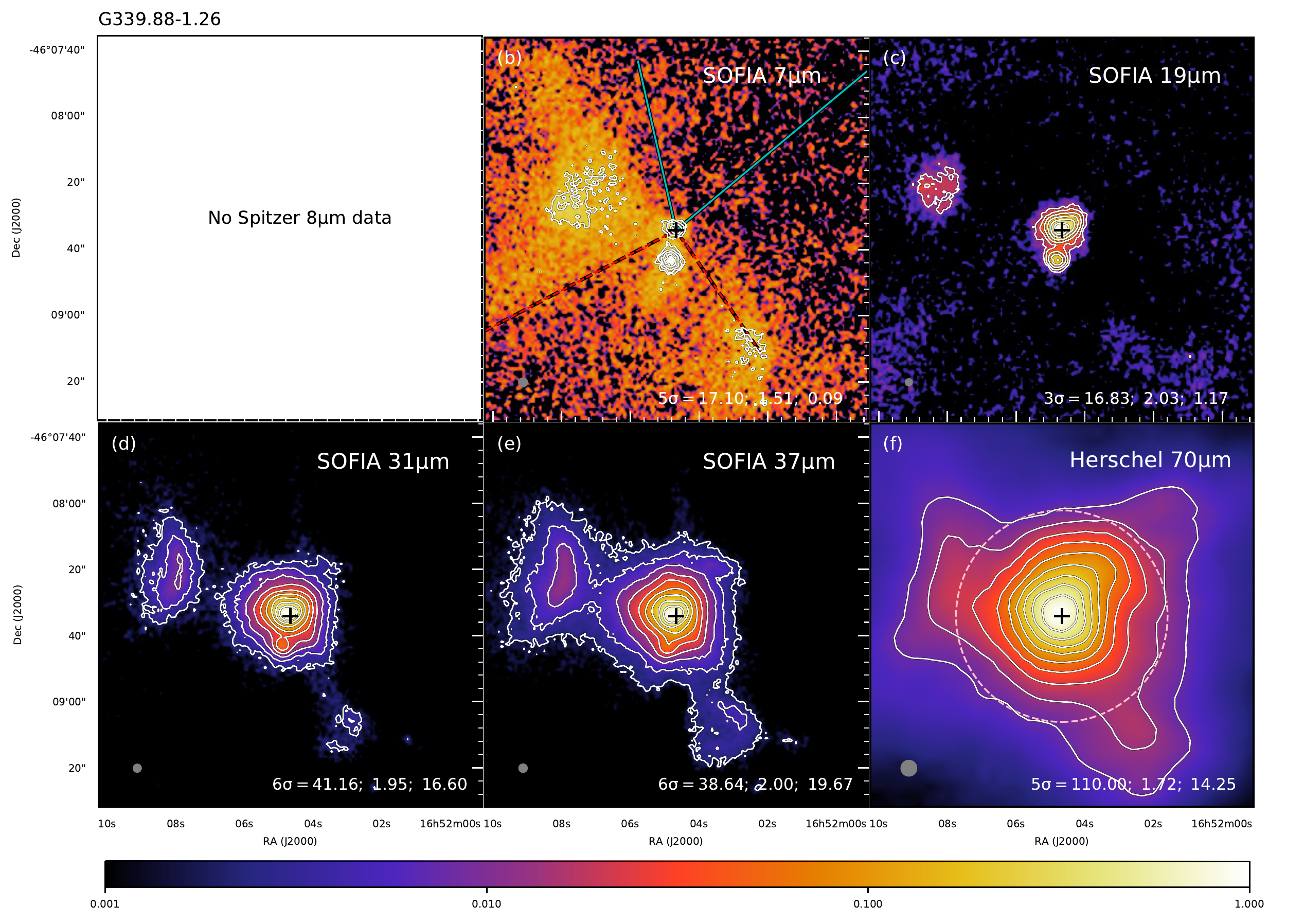}
\caption{
Multi-wavelength images of G339.88-1.26, following the format of
Figure 1. The black cross in all panels denotes the 9\,GHz radio peak
position of the component C from Purser et al. (2016) at R.A.(J2000) =
16$^h$52$^m$04$\fs$67, Decl.(J2000) =
$-$46$\arcdeg$08$\arcmin$34$\farcs$16. \label{fig:G339}}
\end{figure*}

\begin{figure*}
\epsscale{1.1}
\plotone{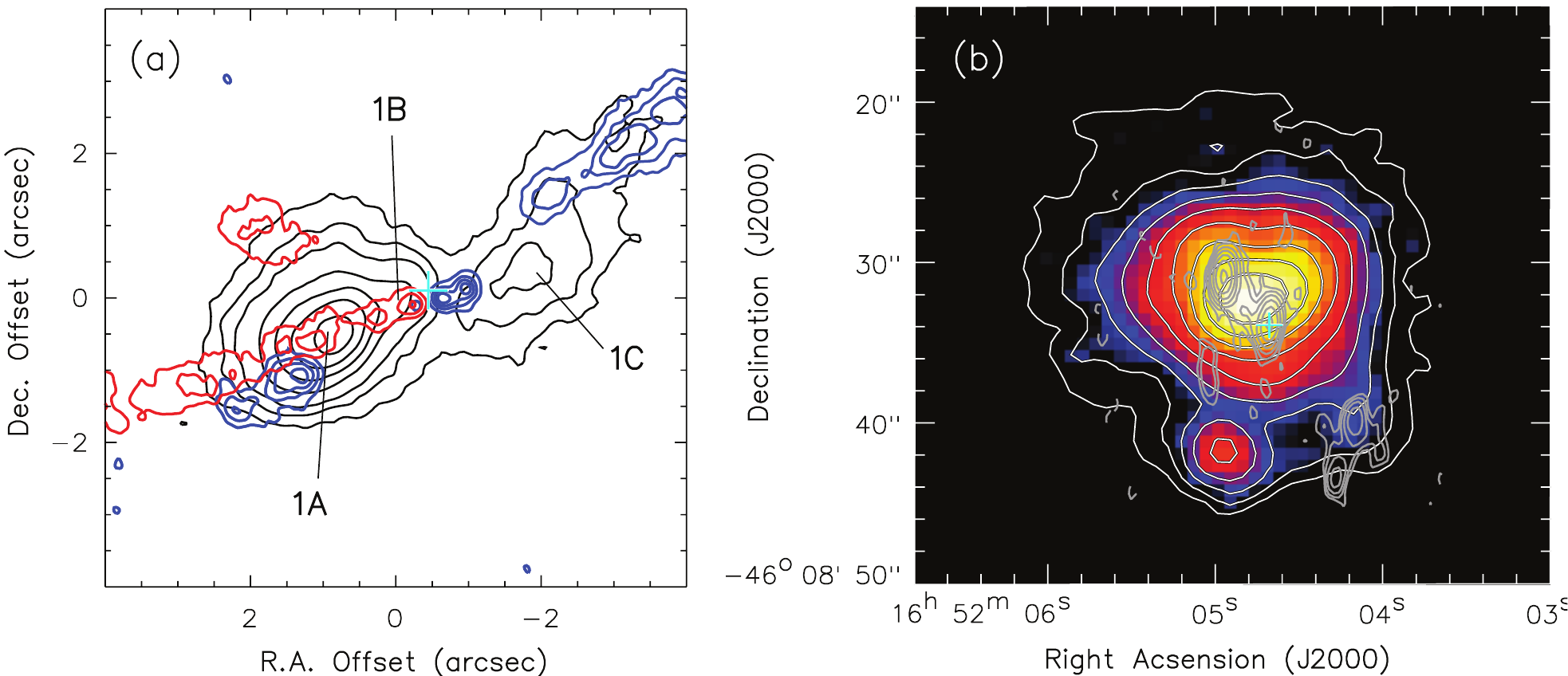}
\caption{
G339.88-1.26. (a) The black contours are the Keck 18\,$\mu$m data,
with the MIR peaks labeled. The red and blue contours show the blue- ($V_{\rm lsr}=-80$ to $-20 \rm \ km\,s^{-1}$) and
red-shifted ($V_{\rm lsr}=-50$ to $+10 \rm \ km\,s^{-1}$) ALMA $^{12}$CO(2--1) observations (systematic velocity $V_{\rm lsr}=-33\rm \ km\,s^{-1}$) by Zhang et
al. (2019). Note that emission from the secondary $^{12}$CO(2--1) outflow is outside the field of view.
The cyan plus sign shows the location of the 8.6\,GHz radio continuum
peak (Ellingsen et al. 1996). (b) The {\it SOFIA} 31\,$\mu$m image in
color and white contours with the 9 GHz radio continuum contours from
Purser et al. (2016). The central radio source is identified as a
radio jet and the two other sources as radio outflow lobes (Purser et
al. 2016). The cyan plus sign shows the location of the 8.6\,GHz radio
continuum peak (Ellingsen et al. 1996). \label{fig:G339_Gemini}}
\end{figure*}

This source, also named IRAS 16484-4603 is located at
$2.1^{+0.4}_{-0.3}$ kpc, determined from trigonometric parallax
measurements of the 6.7 GHz methanol masers using the Australian Long
Baseline Array (Krishnan et al. 2015).

De Buizer et al. (2002) resolved the central MIR emission of G339.88
into three peaks (1A, 1B, and 1C) at 10 and 18\,$\mu$m that all lie
within an extended MIR region elongated at a position angle of
$\sim$120$\arcdeg$ (Figure~\ref{fig:G339_Gemini}a). Interferometric
radio continuum observations have revealed an elongated, ionized
jet/outflow at a position angle of $\sim$45$\arcdeg$ with a scale of
15\arcsec, approximately perpendicular to the elongation of the
infrared emission (Ellingsen et al. 1996; Purser et al. 2016). Recent
ALMA $^{12}$CO(2--1) observations by Zhang et al. (2019) also reveal a
major molecular outflow with a E-W orientation and a tentative second
outflow with a NE-SW orientation (at the same angle as the ionized
outflow seen by Purser et al. 2016). Zhang et al. (2019) suggest that
the 1.3\,mm continuum peak, which is $\sim$ 0.5\arcsec \ to the west of
1B, is the likely location of the origin of both outflows, which may
indicate an unresolved proto-binary system. All of the 10 and
18\,$\mu$m MIR emission seen by De Buizer et al. (2002) is therefore
mainly tracing the outflow cavities of the molecular outflow seen at a
position angle of $\sim$120$\arcdeg$
(Figure~\ref{fig:G339_Gemini}a). Our \textit{SOFIA} data (see
Figure~\ref{fig:G339}) show an extension in this direction as well,
seen best at 19.7\,$\mu$m. At wavelengths longer than 20\,$\mu$m,
there is a faint pull of emission to the NE and another faint
extension to the SW, both of which correspond to the radio lobes of
the ionized outflow (Figure~\ref{fig:G339_Gemini}b). Therefore, both
outflows are revealed in the IR, with the ionized outflow only showing
up at longer wavelengths, which again may be due to
extinction. Detection of red and blue-shifted emission on both sides
suggests a near side-on view of the outflows.

There is a large half-moon feature to the east of the main MIR peak in
our \textit{SOFIA} data. It has radio continuum emission (see
Ellingsen et al. 2005) and could be a cometary compact HII
region. Closer to the main MIR peak, we also see a secondary source
$\sim$ 10\arcsec to the south. There is no CO outflow associated with
this source. We see a counterpart of this source in H and K band as
seen in Figure~\ref{fig:nir}. It could be a more evolved low-mass
protostar. The source that is further SW, which is getting stronger
at wavelengths longer than 31\,$\mu$m, might be related to the ionized
radio jet (Purser et al. 2016), though there is no hint of ionized
emission from it in the study of Ellingsen et al. (2005).

\subsection{General Results from the SOFIA Imaging}

Overall in the sample of sources we have studied here, we often see
that the MIR morphologies appear to be influenced by the presence of
outflow cavities, which create regions of low dust extinction, and the
presence of relatively cool, dense gas structures (potentially
including disks and infall envelopes), which have high dust
extinction, even at relatively long wavelengths. The presence of such
structures is a general feature of Core Accretion models. A number of
sources also appear to have companions, including from being in
regions where a star cluster is likely forming, which can also
complicate the appearance in the MIR.

In addition to the monochromatic images presented above, we also
construct three-color images of all the sources, presented together in
Figure~\ref{fig:rgb}. Note, however, that these RGB images have
different beam sizes for the different colors (especially blue), with
the effect being to tend to give small sources an extended red
halo. In spite of this effect, in G45.12, G309.92, G35.58, IRAS 16562
and G339.88, short wavelength emission seems to dominate the extended
structure. In IRAS 16562, we can see the near-facing outflow cavity
appears bluer while the more extincted, far-facing outflow cavity
appears redder. For the other sources we do not see obvious color
gradients across the sources.

We summarize the properties of the protostellar sources in
Table~\ref{tab:property}. The ordering of the sources is from high to
low for the luminosity estimate (top to bottom). For two out of the three sources with
detected outflows, the MIR morphology is significantly influenced by
outflow cavities. For those lacking outflow information, we consider
that it is still likely that the MIR emission is tracing outflows or
flared disks. Especially in G309.92 and G305.20, high-resolution {\it
  Gemini} data reveals a flat dark lane, which could be the
optically thick disk.

We see that at wavelengths $\ga$\,19\,$\mu$m, there is an offset
between the radio emission, if that is where the protostar is located,
and the MIR peaks in G309.92, G35.58, G49.27, G339.88. In Paper I we
found that the MIR peaks appear displaced away from the protostar
towards the blue-shifted, near-facing side of the outflow due to the
higher extinction of the far-facing side at short wavelength. Here
G339.88 may reveal a hint of this trend of the displacement. For the
other sources, due to the lack of outflow information, the cause of
the offset is not yet clear.

We have also found candidates of more evolved, probably lower-mass
protostars in the company of the massive protostar in most regions,
based on the common peaks seen in multi-wavelength MIR and NIR data
and how their fluxes change with wavelength. With the caveat that
  our sample is likely incomplete, the projected separation between
the massive protostar and the nearest lower-mass companion revealed by {\it SOFIA} is about
0.28\,pc in G45.12, 0.49\,pc in G35.58, 0.12\,pc in IRAS16562,
0.28\,pc in G305.20, and 0.10\,pc in G339.88. Note that Core
  Accretion models, such as the Turbulent Core Model of McKee \& Tan
  (2003), can be applied to conditions inside protoclusters, as well
  as to more isolated regions, while Competitive Accretion (Bonnell et
  al. 2001) and Protostellar Collision (Bonnell et al. 1998) models
  require the presence of a rich stellar cluster around the
  protostar. To the extent that some of the presented sources appear
  to be in relatively isolated environments is thus tentative evidence
  in support of Core Accretion models, but deeper observations to
  probe the low-mass stellar population are needed to confirm this.

\begin{figure*}
\epsscale{1.17}
\plotone{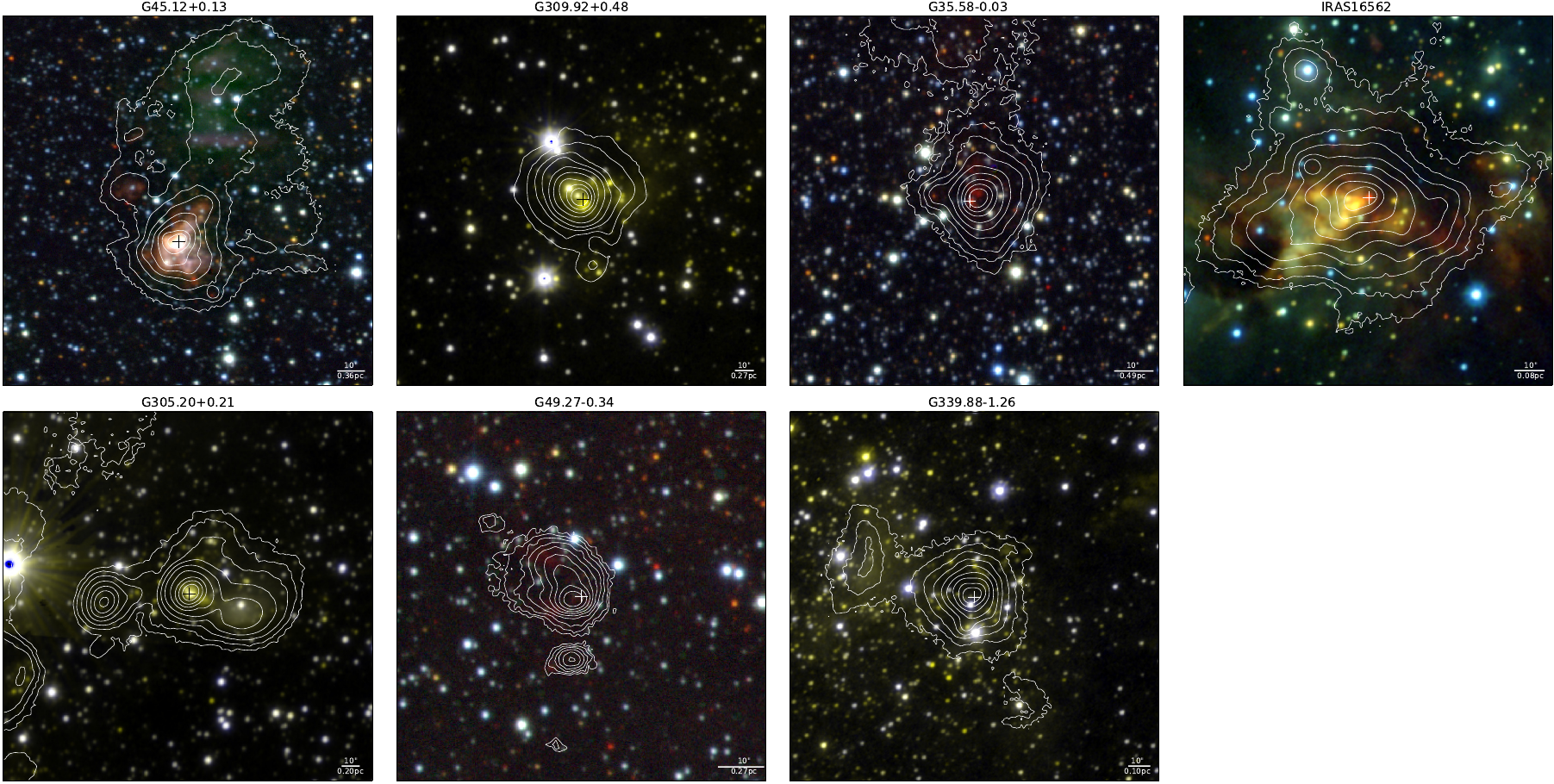}
\caption{
NIR RGB images of the seven protostellar sources, as labeled. The data
of G45.12, G35.58, G49.27 and S235 come from the UKIDSS survey. The
data of G309, IRAS16562, G305 and G339 come from the VVV survey.  K
band data is shown in red. H band data is shown in green. J band data
is shown in blue. The white contours are SOFIA 37$\mu$m emission, with
the same levels displayed in the previous individual figures for each
source. The crosses in each panel are the same as the crosses in the
previous individual figures, denoting the radio sources (methanol
maser in G305). The scale bar is shown in the right corner of each
panel.}\label{fig:nir}
\end{figure*}

\begin{figure*}
\epsscale{1.17}
\plotone{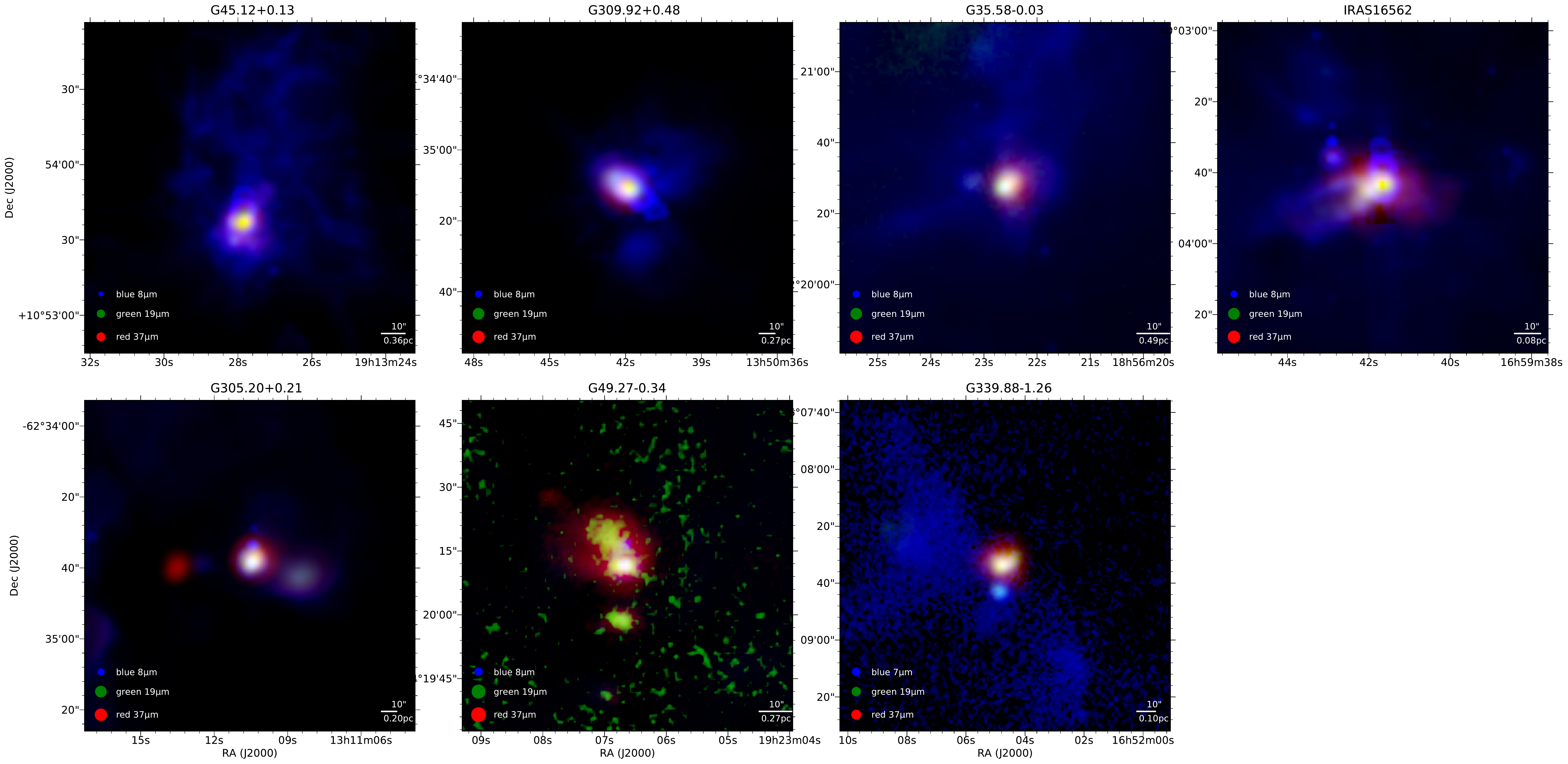}
\caption{
Gallery of RGB images of the seven protostellar sources, as
labeled. The color intensity scales are stretched as arcsinh and show
a dynamic range of 100 from the peak emission at each wavelength,
except for the 19\,$\mu$m image of G49.27, where only a dynamic range
of 10 is shown due to its relatively low signal to noise ratio. The
legend shows the wavelengths used and the beam sizes at these
wavelengths. {\it SOFIA}-FORCAST 37\,$\mu$m is shown in
red. SOFIA-FORCAST 19\,$\mu$m is shown in green. Blue usually shows
{\it Spitzer} IRAC 8 $\mu$m, except for G339.88-1.26, where it
displays {\it SOFIA}-FORCAST 7 $\mu$m.  \label{fig:rgb}}
\end{figure*}

\clearpage
\begin{deluxetable*}{ccccc}
\setlength{\tabcolsep}{4pt}
%\tabletypesize{\scriptsize}
\tablecaption{Summary of Properties of the Protostellar Sources \label{tab:property}}
\tablewidth{18pt}
\tablehead{
\colhead{Source} & \colhead{Radio emission?} & \colhead{Outflow?} & \colhead{Multiple (proto)stars within 20\arcsec?} & \colhead{What regulates the MIR morphology?} 
}
\startdata
G45.12+0.13 & UC HII & Two & Cluster\tablenotemark{a}. & Ionized medium \\
\hline\noalign{\smallskip}
\multirow{2}{*}{G309.92+0.48} & \multirow{2}{*}{HC HII}  & \multirow{2}{*}{...} & MIR companion. & Outflow cavities \\
&  &  & Resolved. & or flared disk surface? \\
\hline\noalign{\smallskip}
\multirow{2}{*}{G35.58-0.03} & \multirow{2}{*}{UC HII} & \multirow{2}{*}{N} & Nearby HII region with an outflow. & Outflows from nearby sources?\\
&  &  &  Low-mass YSO\tablenotemark{b}? & \\
\hline\noalign{\smallskip}
IRAS 16562-3959 & HC HII with jet & Two & Cluster\tablenotemark{c}. & Outflow cavities \\
\hline\noalign{\smallskip}
\multirow{2}{*}{G305.20+0.21} & \multirow{2}{*}{N} & \multirow{2}{*}{...} & Nearby HII region. MIR companion. & Outflow cavities \\
 &  &  & Resolved\tablenotemark{d}. Low-mass YSO\tablenotemark{b}? & or flared disk surface? \\
 \hline\noalign{\smallskip}
G49.27-0.34 & Y & ... & Radio companion. MIR companion. & ... \\
\hline\noalign{\smallskip}
\multirow{2}{*}{G339.88-1.26} & \multirow{2}{*}{Jet} & \multirow{2}{*}{Two} & MIR companion. Resolved\tablenotemark{e}. Binary\tablenotemark{f}? & Outflow cavities  \\
 & &  & Low-mass YSO\tablenotemark{b}?  Nearby HII region? & and extinction \\
\enddata
\tablenotetext{a}{Based on radio sources from Vig et al. 2006.}
\tablenotetext{b}{Based on multi-wavelength MIR and NIR data.}
\tablenotetext{c}{Based on X-ray sources from Montes et al. 2018.}
\tablenotetext{d}{We suspect here the resolved structures are more likely to be emission separated by optically thick disk rather than two distinct protostars.}
\tablenotetext{e}{We suspect here the resolved structures are emission tracing the outflow cavities rather than multiple distinct protostars. See also Zhang et al. 2019.}
\tablenotetext{f}{Based on the fact of two outflows.}
\end{deluxetable*}
 
\begin{figure*}
\epsscale{1.2}
\plotone{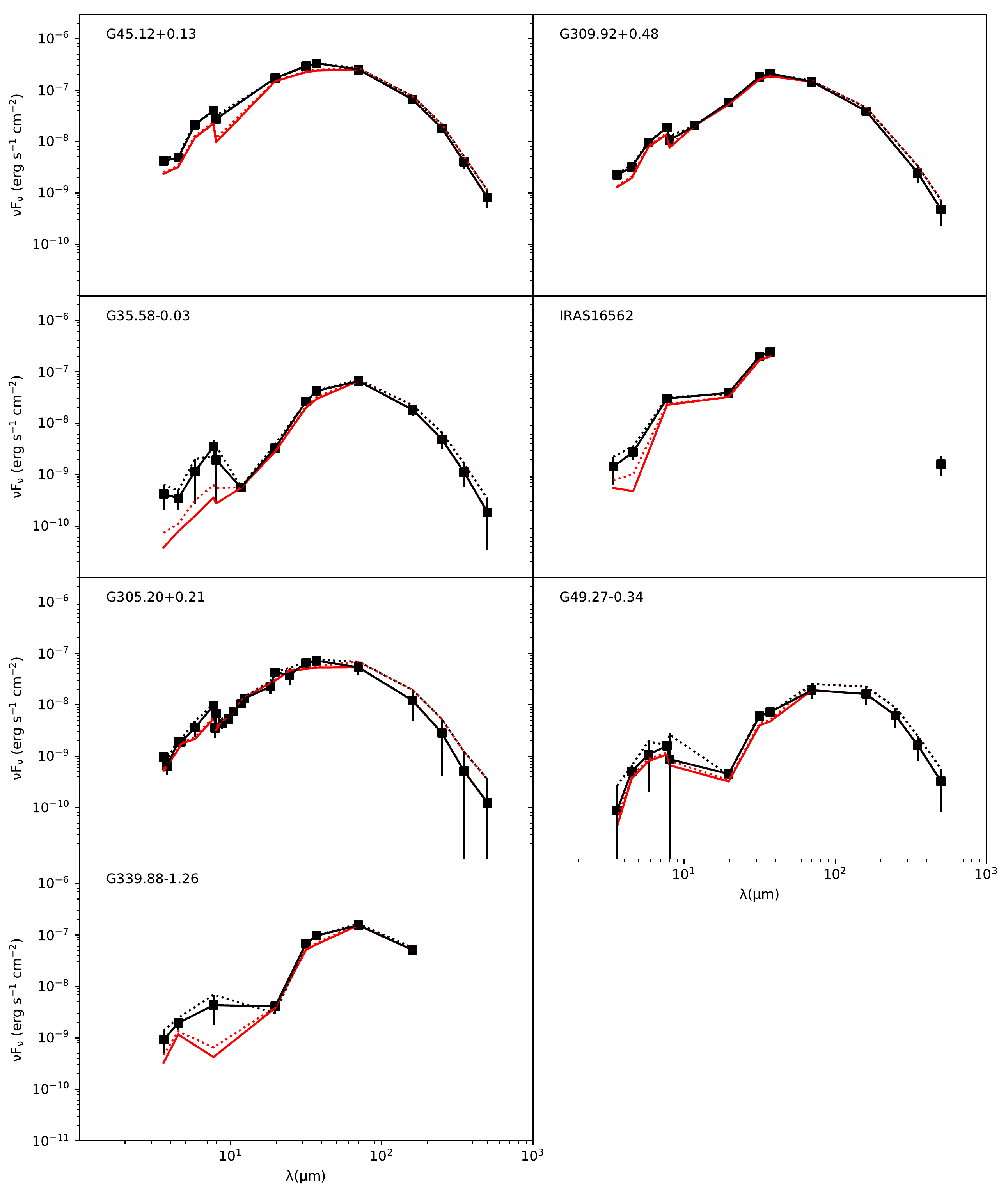}
%\vspace{-0.5in}
\caption{
SEDs of the seven presented sources. Total fluxes with no background
subtraction applied are shown by dotted lines. The fixed aperture case
is black dotted; the variable aperture (at $<70\:{\rm \mu m}$) case is
red dotted. The background subtracted SEDs are shown by solid lines:
black for fixed aperture (the fiducial case); red for variable
aperture. Black solid squares indicate the actual measured values that
sample the fiducial SED. Note the Spitzer 4.5 $\mu$m, 5.8 $\mu$m and 8
$\mu$m data of G309 and all Spitzer data of G45.12
have ghosting problems and are not used for the SED fitting.
%Flux densities are derived with background subtraction in fixed
%apertures (black solid, fiducial SED), without background subtraction
%in fixed apertures (black dotted), with background subtraction in
%various apertures (red solid) and without background subtraction in
%various apertures (red dotted). 
%The black squares denote data points with background subtraction in
%fixed apertures. 
%The empty squares in G35.20-0.74 denote data points
%without background subtraction in fixed apertures. 
\label{fig:SEDs}}
\end{figure*}

\begin{figure*}
\epsscale{1.10}
\plotone{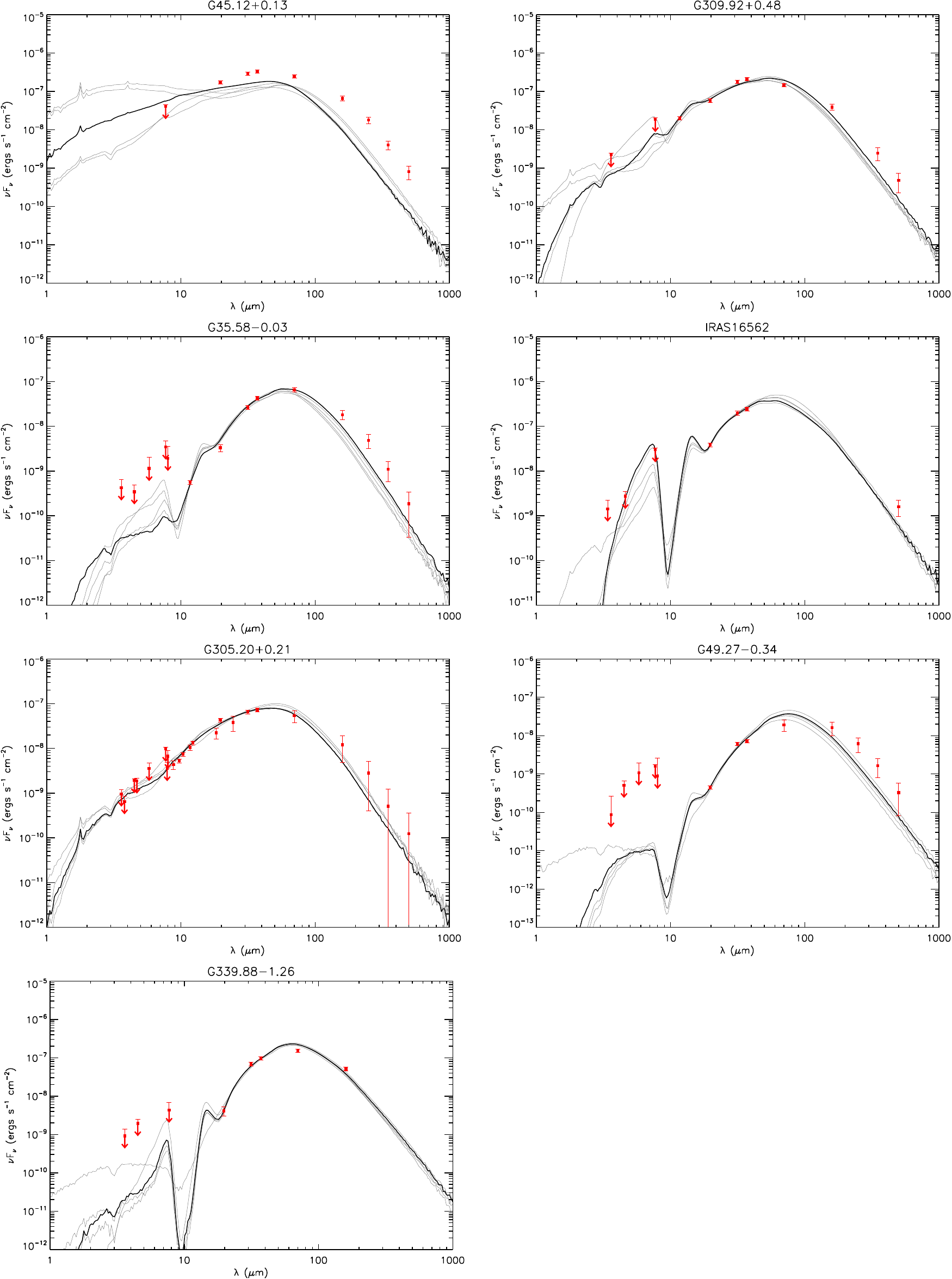}
%\vspace{-3cm}
\caption{
Protostar model fitting to the fixed aperture, background-subtracted
SED data using the ZT model grid. For each source, the best fit model
is shown with a solid black line and the next four best models are
shown with solid gray lines. Flux values are those from
Table~\ref{tab:flux}. Note that the data at $\lesssim8\:{\rm \mu m}$
are treated as upper limits (see text). The resulting model parameter
results are listed in Table~\ref{tab:models}.\label{fig:SEDsZT}}
\end{figure*}

\clearpage
\renewcommand{\arraystretch}{0.9}
\begin{deluxetable*}{ccccccccccccc}
\tabletypesize{\scriptsize}
\tablecaption{Parameters of the Best Five Fitted Models \label{tab:models}} 
\tablewidth{18pt}
\tablehead{
\colhead{Source} &\colhead{$\chi^{2}$/N} & \colhead{$M_{\rm c}$} & \colhead{$\Sigma_{\rm cl}$} & \colhead{$R_{\rm core}$}  &\colhead{$m_{*}$} & \colhead{$\theta_{\rm view}$} &\colhead{$A_{V}$} & \colhead{$M_{\rm env}$} &\colhead{$\theta_{w,\rm esc}$} & \colhead{$\dot {M}_{\rm disk}$} & \colhead{$L_{\rm bol, iso}$} & \colhead{$L_{\rm bol}$} \\
\colhead{} & \colhead{} & \colhead{($M_\odot$)} & \colhead{(g $\rm cm^{-2}$)} & \colhead{(pc) ($\arcsec$)} & \colhead{($M_{\odot}$)} & \colhead{(\arcdeg)} & \colhead{(mag)} & \colhead{($M_{\odot}$)} & \colhead{(deg)} &\colhead{($M_{\odot}$/yr)} & \colhead{($L_{\odot}$)} & \colhead{($L_{\odot}$)} \\
 \vspace{-0.4cm}
}
\startdata
G45.12+0.13
& 54.39 & 480 & 1.0 & 0.161 ( 4 ) & 64.0 & 34 & 0.0 & 325 & 32 & 1.2$\times 10^{-3}$ & 6.5$\times 10^{5}$ & 8.4$\times 10^{5}$ \\
$d$ = 7.4 kpc
& 63.23 & 480 & 1.0 & 0.161 ( 4 ) & 48.0 & 29 & 0.0 & 367 & 25 & 1.1$\times 10^{-3}$ & 4.5$\times 10^{5}$ & 5.4$\times 10^{5}$ \\
$R_{ap}$ = 48 \arcsec
& 65.40 & 480 & 3.2 & 0.091 ( 3 ) & 24.0 & 13 & 0.0 & 441 & 12 & 2.0$\times 10^{-3}$ & 1.1$\times 10^{6}$ & 2.9$\times 10^{5}$ \\
= 1.72 pc
& 66.41 & 400 & 3.2 & 0.083 ( 2 ) & 24.0 & 13 & 0.0 & 362 & 13 & 1.9$\times 10^{-3}$ & 1.3$\times 10^{6}$ & 3.0$\times 10^{5}$ \\
& 69.30 & 240 & 3.2 & 0.064 ( 2 ) & 32.0 & 29 & 0.0 & 175 & 23 & 1.9$\times 10^{-3}$ & 4.5$\times 10^{5}$ & 5.0$\times 10^{5}$ \\
\hline\noalign{\smallskip}
G309.92+0.48
& 2.82 & 320 & 3.2 & 0.074 ( 3 ) & 24.0 & 22 & 12.1 & 277 & 15 & 1.8$\times 10^{-3}$ & 3.3$\times 10^{5}$ & 3.1$\times 10^{5}$ \\
$d$ = 5.5 kpc
& 3.90 & 480 & 1.0 & 0.161 ( 6 ) & 48.0 & 29 & 39.4 & 367 & 25 & 1.1$\times 10^{-3}$ & 4.5$\times 10^{5}$ & 5.4$\times 10^{5}$ \\
$R_{ap}$ = 32 \arcsec
& 4.38 & 240 & 3.2 & 0.064 ( 2 ) & 32.0 & 34 & 17.2 & 175 & 23 & 1.9$\times 10^{-3}$ & 3.2$\times 10^{5}$ & 5.0$\times 10^{5}$ \\
= 0.85 pc
& 4.71 & 240 & 3.2 & 0.064 ( 2 ) & 24.0 & 29 & 0.0 & 194 & 18 & 1.6$\times 10^{-3}$ & 2.6$\times 10^{5}$ & 3.1$\times 10^{5}$ \\
& 4.97 & 400 & 1.0 & 0.147 ( 6 ) & 48.0 & 34 & 4.0 & 289 & 29 & 1.0$\times 10^{-3}$ & 3.0$\times 10^{5}$ & 5.3$\times 10^{5}$ \\
\hline\noalign{\smallskip}
G35.58-0.03
& 1.70 & 480 & 3.2 & 0.091 ( 2 ) & 24.0 & 22 & 16.2 & 441 & 12 & 2.0$\times 10^{-3}$ & 2.9$\times 10^{5}$ & 2.9$\times 10^{5}$ \\
$d$ = 10.2 kpc
& 2.14 & 400 & 3.2 & 0.083 ( 2 ) & 24.0 & 22 & 46.5 & 362 & 13 & 1.9$\times 10^{-3}$ & 3.0$\times 10^{5}$ & 3.0$\times 10^{5}$ \\
$R_{ap}$ = 26 \arcsec
& 3.41 & 320 & 3.2 & 0.074 ( 1 ) & 24.0 & 29 & 35.4 & 277 & 15 & 1.8$\times 10^{-3}$ & 2.7$\times 10^{5}$ & 3.1$\times 10^{5}$ \\
= 1.27 pc
& 4.28 & 480 & 1.0 & 0.161 ( 3 ) & 48.0 & 34 & 39.4 & 367 & 25 & 1.1$\times 10^{-3}$ & 3.0$\times 10^{5}$ & 5.4$\times 10^{5}$ \\
& 4.49 & 480 & 1.0 & 0.161 ( 3 ) & 64.0 & 39 & 72.7 & 325 & 32 & 1.2$\times 10^{-3}$ & 3.6$\times 10^{5}$ & 8.4$\times 10^{5}$ \\
\hline\noalign{\smallskip}
IRAS16562
& 0.53 & 400 & 0.1 & 0.465 ( 56 ) & 32.0 & 44 & 100.0 & 304 & 29 & 1.5$\times 10^{-4}$ & 9.2$\times 10^{4}$ & 1.6$\times 10^{5}$ \\
$d$ = 1.7 kpc
& 0.64 & 480 & 0.1 & 0.510 ( 62 ) & 24.0 & 71 & 55.6 & 418 & 21 & 1.4$\times 10^{-4}$ & 5.7$\times 10^{4}$ & 8.7$\times 10^{4}$ \\
$R_{ap}$ = 32 \arcsec
& 0.65 & 480 & 0.1 & 0.510 ( 62 ) & 32.0 & 48 & 100.0 & 391 & 26 & 1.6$\times 10^{-4}$ & 9.8$\times 10^{4}$ & 1.6$\times 10^{5}$ \\
= 0.26 pc
& 0.67 & 320 & 0.3 & 0.234 ( 28 ) & 16.0 & 22 & 17.2 & 283 & 16 & 2.5$\times 10^{-4}$ & 5.3$\times 10^{4}$ & 6.1$\times 10^{4}$ \\
& 0.83 & 120 & 3.2 & 0.045 ( 6 ) & 16.0 & 29 & 100.0 & 90 & 21 & 1.1$\times 10^{-3}$ & 1.0$\times 10^{5}$ & 1.2$\times 10^{5}$ \\
\hline\noalign{\smallskip}
G305.20+0.21
& 0.79 & 80 & 3.2 & 0.037 ( 2 ) & 24.0 & 48 & 14.1 & 35 & 37 & 1.1$\times 10^{-3}$ & 7.5$\times 10^{4}$ & 2.6$\times 10^{5}$ \\
$d$ = 4.1 kpc
& 0.92 & 100 & 3.2 & 0.041 ( 2 ) & 32.0 & 51 & 18.2 & 37 & 42 & 1.2$\times 10^{-3}$ & 7.9$\times 10^{4}$ & 3.5$\times 10^{5}$ \\
$R_{ap}$ = 16 \arcsec
& 0.97 & 160 & 1.0 & 0.093 ( 5 ) & 32.0 & 44 & 13.1 & 88 & 39 & 5.9$\times 10^{-4}$ & 8.2$\times 10^{4}$ & 2.3$\times 10^{5}$ \\
= 0.32 pc
& 1.04 & 80 & 3.2 & 0.037 ( 2 ) & 16.0 & 34 & 8.1 & 50 & 27 & 9.5$\times 10^{-4}$ & 7.2$\times 10^{4}$ & 1.1$\times 10^{5}$ \\
& 1.11 & 160 & 3.2 & 0.052 ( 3 ) & 48.0 & 58 & 16.2 & 59 & 45 & 1.6$\times 10^{-3}$ & 9.0$\times 10^{4}$ & 6.4$\times 10^{5}$ \\
\hline\noalign{\smallskip}
G49.27-0.34
& 1.87 & 240 & 3.2 & 0.064 ( 2 ) & 12.0 & 22 & 54.5 & 219 & 12 & 1.2$\times 10^{-3}$ & 4.5$\times 10^{4}$ & 4.8$\times 10^{4}$ \\
$d$ = 5.55 kpc
& 1.96 & 200 & 3.2 & 0.059 ( 2 ) & 12.0 & 22 & 92.9 & 179 & 13 & 1.1$\times 10^{-3}$ & 4.9$\times 10^{4}$ & 5.2$\times 10^{4}$ \\
$R_{ap}$ = 29 \arcsec
& 2.18 & 320 & 3.2 & 0.074 ( 3 ) & 12.0 & 22 & 0.0 & 302 & 10 & 1.3$\times 10^{-3}$ & 4.7$\times 10^{4}$ & 4.9$\times 10^{4}$ \\
= 0.77 pc
& 2.37 & 160 & 3.2 & 0.052 ( 2 ) & 12.0 & 29 & 77.8 & 139 & 15 & 1.0$\times 10^{-3}$ & 4.4$\times 10^{4}$ & 5.3$\times 10^{4}$ \\
& 2.73 & 120 & 3.2 & 0.045 ( 2 ) & 12.0 & 34 & 73.7 & 99 & 18 & 9.6$\times 10^{-4}$ & 3.6$\times 10^{4}$ & 5.2$\times 10^{4}$ \\
\hline\noalign{\smallskip}
G339.88-1.26
& 2.21 & 400 & 0.3 & 0.262 ( 26 ) & 12.0 & 22 & 17.2 & 373 & 11 & 2.3$\times 10^{-4}$ & 3.7$\times 10^{4}$ & 4.0$\times 10^{4}$ \\
$d$ = 2.1 kpc
& 2.30 & 320 & 0.3 & 0.234 ( 23 ) & 12.0 & 68 & 6.1 & 293 & 13 & 2.2$\times 10^{-4}$ & 3.3$\times 10^{4}$ & 4.0$\times 10^{4}$ \\
$R_{ap}$ = 32 \arcsec
& 2.48 & 480 & 0.3 & 0.287 ( 28 ) & 12.0 & 22 & 7.1 & 459 & 10 & 2.5$\times 10^{-4}$ & 3.8$\times 10^{4}$ & 4.0$\times 10^{4}$ \\
= 0.33 pc
& 2.62 & 320 & 0.3 & 0.234 ( 23 ) & 16.0 & 22 & 90.9 & 283 & 16 & 2.5$\times 10^{-4}$ & 5.3$\times 10^{4}$ & 6.1$\times 10^{4}$ \\
& 2.84 & 120 & 3.2 & 0.045 ( 4 ) & 12.0 & 44 & 0.0 & 99 & 18 & 9.6$\times 10^{-4}$ & 3.3$\times 10^{4}$ & 5.2$\times 10^{4}$ \\
\enddata
\end{deluxetable*}

\subsection{Results of SED Model Fitting}

\subsubsection{The SEDs}\label{S:SED results}

Figure~\ref{fig:SEDs} shows the SEDs of the seven sources that have
been discussed in this paper. The figure illustrates the effects of
using fixed or variable apertures, as well as the effect of background
subtraction. Our fiducial method is that with fixed aperture and with
background subtraction carried out. This tends to have moderately
larger fluxes at shorter wavelengths than the variable aperture SED
especially for G35.58, IRAS16562 and G339.88 where emission from
secondary sources can be significant at wavelengths $\leq$
8\,$\mu$m. However, as in Paper I, the $\leq$ 8\,$\mu$m flux is in any
case treated as an upper limit in the SED model fitting, given the
difficulties of modeling emission from PAHs and transiently heated
small grains. The flux density derived from the two methods between
10\,$\mu$m and 70\,$\mu$m is generally close. For flux densities
longer than 70\,$\mu$m, the influence of secondary sources is not
illustrated by the variable aperture method. However, we tried
measuring the SEDs up to 37\,$\mu$m of the MIR companions alone, which
are resolved from the emission of the main protostar, and found that
their flux density at each wavelength is $\leq$ 5\% of that of the
main protostar (except that the 19\,$\mu$m flux density of the
southern patch in G49.27 is $\sim$ 20\% of that of the massive
protostar). Moreover, all of them have a SED peak $\leq$ 31\,$\mu$m
except that the southern patch in G49.27 has a nearly flat rising
slope between 31\,$\mu$m and 37\,$\mu$m. Thus the influence of
secondary sources is generally not severe at long wavelengths that
control the SED fitting.

%well included in the discrepancy between the SEDs derived from the two
%methods with the influence $\geq$ 70\,$\mu$m being minimal.

Again, as mentioned in \S\ref{S:SED construction}, for the cases where
there seem to be multiple sources in the fiducial aperture, the model
assumes one source dominates the luminosity and the key is to measure
the flux from the same region across all wavelengths. If a source is
isolated, then the fixed aperture at shorter wavelengths, which tends to
be larger than the source appears, may include more noise and make the
photometry less accurate than the variable aperture
method. However, since we set the clump background emission as
  the magnitude of the uncertainty, this effect should be very minor.

The peaks of the SEDs are generally between 37\,$\mu$m and
70\,$\mu$m. In particular, the SED peaks of G45.12, G309.92, G305.20
appear to be closer to 37\,$\mu$m, while the peaks of G35.58, G49.27
and G339.88 appear to be closer to 70\,$\mu$m. This may be related to
the evolutionary stage and/or viewing angle of the sources (see
\S\ref{S:fitting results}).

\subsubsection{ZT Model Fitting Results}\label{S:fitting results}

Figure~\ref{fig:SEDsZT} shows the results of fitting the ZT
protostellar radiative transfer models to the fixed aperture,
background-subtracted SEDs. Note that the data at $\leq$ 8 $\mu$m are
considered to be upper limits given that PAH emission and transiently
heated small grain emission are not well treated in the models. The
parameters of the best-fit ZT models are listed in
Table~\ref{tab:models}. From left to right the parameters are
reduced $\chi^{2}$ (i.e., normalized by the number of data points in
the SED, $N$), the initial core mass ($M_{c}$), the mean mass
surface density of the clump ($\Sigma_{\rm cl}$), the initial core
radius ($R_{\rm core}$), the current protostellar mass ($m_{*}$), the
viewing angle ($\theta_{\rm view}$), foreground extinction ($A_{V}$),
current envelope mass ($M_{\rm env}$), half opening angle of the
outflow cavity ($\theta_{w,\rm esc}$), accretion rate from the disk to
the protostar ($\dot{m}_{*}$),
%jct - note above notation change... I think this is better than \dot{M}_disk... check with Yichen
the luminosity integrated from the unextincted model SEDs assuming
isotropic radiation ($L_{\rm bol, iso}$), and the inclination
corrected, true bolometric luminosity ($L_{\rm bol}$).  For each
source, the best five models are shown, ordered from best to worst as
measured by $\chi^2$. Note that these are distinct physical models
with differing values of $M_{c}$, $\Sigma_{\rm cl}$, and/or $m_{\rm
  *}$, i.e., we do not display simple variations of $\theta_{\rm
  view}$ or $A_{V}$ for each of these different physical models.

The best-fit models imply the sources have protostellar masses $m_{*}
\sim 12-64 \: M_{\odot}$ accreting at rates of $\dot{m}_{*} \sim
10^{-4}-10^{-3} \: M_{\odot} \: \rm yr^{-1}$ inside cores of initial
masses $M_{c} \sim 100-500 \: M_{\odot}$ embedded in clumps with mass
surface densities $\Sigma_{\rm cl} \sim 0.1-3 \: \rm g \ cm^{-2}$ and
span a luminosity range of $10^{4} -10^{6} \: L_{\odot}$.

In most sources the best five models have similar values of $\chi^2$,
but there is still significant variations in the model parameters even
for G305.20 which has the most SED data points. As stated in Paper I,
this illustrates the degeneracy in trying to constrain the
protostellar properties from only MIR to FIR SEDs, which would be improved by extended SEDs fitting including centimeter continuum flux densities (Rosero et al. 2019) and image intensity profile comparison. 
From the SED shape
the most variation between models appears at shorter wavelengths. Here
more data points can help better constrain the models, as in
G305.20. Again we note that although sometimes the $\chi^2$ may look
high, as in G45.12+0.13, here we focus more on the relative comparison
of $\chi^2$ between the models available in the model grid, which
still give us constraints on the protostellar properties. At
wavelengths $> \ 70 \ \mu$m the models tend to be lower than the data
points in many sources. Note the values of $R_{\rm core}$ returned by
the models are usually much smaller than the aperture radii. This
would indicate that, even after a first attempt at clump
  background subtraction, the measured flux still has significant 
  contribution from the cool surrounding clump. Recall that this
  component is not included in the ZT radiative transfer models and
  can thus lead to the offset at long wavelengths, i.e., with models
  under-predicting the observed fluxes.

We also tried fitting the SEDs with variable apertures across
wavelength. Most sources have $R_{\rm core}$ similar to that derived
in the fiducial case and still the models appear lower than the data
points at long wavelengths for G309.92, G35.58, G305.20 and G49.27.
%Without setting IRAC as upper limits: G49.27 returns $R_{\rm core}$ $\sim$ 20 \arcsec with larger $M_{\rm c}$ and smaller $\Sigma_{\rm cl}$, $m_{\rm *}$. But the models appear even lower at wavelengths $>$ 70 $\mu$m. G339 returns $R_{\rm core}$ $\sim$ 10 \arcsec with smaller $M_{\rm c}$ and larger $\Sigma_{\rm cl}$, $m_{\rm *}$. 
%Setting upper limits: G49.27 two models return $R_{\rm core}$ $\sim$ 20 \arcsec. G339 three models return $R_{\rm core}$ $\sim$ 5 \arcsec with a smaller $M_{\rm c}$.
 
We note that $m_{*}$ appears quite high for G45.12+0.13, G309.92+0.48,
G35.58-0.03, IRAS 16562, G305.20+0.21. This is likely due to there
being more than one protostar inside the aperture, even with the
variable aperture case, like the source G35.20-0.74 in Paper I, where
the stellar mass returned by the models is around the sum of the two
binary protostars in the center (Beltr\'an et al. 2016, Zhang et
al. 2018).

The location of the SED peak is thought to show a dependence on the
evolutionary stage of the source. We compare the current age derived
from the models and the corresponding total star formation time scale
based on Eq. (44) in MT03 assuming a star formation efficiency of
0.5. G305.20 appears to be the most evolved followed by G309.92 and
G45.12. G339.88 appears to be the least evolved followed by
G49.27. G339.88 is still deep embedded with high dust extinction while
G49.27 is an IRDC source. They are likely the youngest YSOs among the
seven sources. The evolutionary stage revealed by the models is
consistent with the picture that more evolved sources have a SED peak
located at shorter wavelengths, as described in \S\ref{S:SED results}.
However, orientation effects may also be playing a role, since the
peak of the SED shifts to shorter wavelengths when viewing sources at
angles closer to their outflow axis.

Next we describe the fitting results of each individual source and
compare with previous literature results.

\textbf{G45.12+0.13:} This is our most luminous source (almost $10^{6}
\: L_{\odot}$) and hits the boundary of the parameter space of the ZT
model grid, which is partly why the models do not seem to fit the data
points very well, as shown in Figure~\ref{fig:SEDsZT}, since there are only
a few models around $10^{6} \: L_{\odot}$ (Zhang \& Tan 2018). As an
experiment, we tried changing the distance from 7.4 kpc to 1 kpc and
were able to obtain fitting results that have much smaller values of
$\chi^2$. On the other hand, this region is likely to be a
protocluster hosting many ZAMS stars. Thus the assumption of one
source dominating the luminosity may not work well here. The current
best models indicate high initial core mass $M_{c} \sim
500\:M_{\odot}$, high $\Sigma_{\rm cl} \ga 1.0 \rm \ g \: cm^{-2}$
clump environment and high protostellar mass $m_{*} \geq
24\:M_{\odot}$ for the dominant source. The accretion rate is $\sim
10^{-3} M_{\odot} \: {\rm yr}^{-1}$. The current envelope mass is also
typically as high as $\sim$ 400~$M_{\odot}$. The foreground extinction
$A_V$ is estimated to be very low, but this may be an artefact of
other problems of the model fitting.
%This implies the protocluster may form from a even more massive
%environment.
% still embedded in a massive envelope? still accreting?
The best five models all give a $\theta_{\rm view}$ close to
$\theta_{w,\rm esc}$, which leads to high levels of short wavelength
emission.

%jctfinal - commenting out this paragraph - it does not make sense to me, including the strange dust to gas mass ratio... maybe add back later after submission
%A dust mass of 36 $M_{\odot}$ was obtained by Hoare et al. (1991)
%(scaled to 7.4 kpc) from submillimeter observations. Vig et al. (2006)
%carried out simple radiative transfer modeling to fit the SEDs with a
%self-consistent scheme developed by Mookerjea \& Ghosh (1999). The
%dust mass obtained from emission at 450 and 850 microns is 20
%$M_{\odot}$, while that obtained from 200 micron emission map is 18
%$M_{\odot}$. A dust mass of 12 $M_{\odot}$ is obtained from the
%radiative transfer modeling. Considering a gas-to-dust ratio of 250 by
%mass from the radiative transfer model, they determine the gas mass in
%this region to be 5000 $M_{\odot}$.

%jct??? commenting out...
%These compare well with the stellar mass returned in three of our best
%models.

\textbf{G309.92+0.48:}
The best models prefer a massive protostar of $\sim$ 24 to 48
$M_{\odot}$ accreting at $\sim 10^{-3} M_{\odot} \: {\rm yr}^{-1}$ in
a massive core of $\sim$ 240 to 480 $M_{\odot}$ in high $\Sigma_{\rm
  cl} \ga 1.0\ \rm g \ cm^{-2}$ clump environments. The protostar is
slightly inclined $\sim$ 30$\arcdeg$. Walsh et al. (1997) concluded
that if the region were powered by a single star, it would have to be
an O5.5 star with a luminosity of $3.1 \times 10^{5} L_{\odot}$, which
agrees well with the isotropic luminosities returned by our
models. The viewing angle is close to the outflow half opening angle,
resulting in a relatively flat SED shape at shorter wavelengths.

\textbf{G35.58-0.03:}
The best models prefer a massive protostar of $\sim$ 24 to 64
$M_{\odot}$ accreting at $\sim 10^{-3} M_{\odot} \: {\rm yr}^{-1}$ in
a massive core of $\sim$ 320 to 480 $M_{\odot}$ in high $\Sigma_{\rm
  cl} \ga 1.0 \: \rm g \: cm^{-2}$ clump environments. We also tried
fitting the SEDs with the flux measured in variable apertures without
setting short wavelength data as upper limits, which exclude the flux
from the secondary source to the east at short wavelengths. The best
five models have almost the same range for $M_{c}$, $\Sigma_{\rm cl}$,
$m_{*}$, $\dot{m}_{*}$ and $L_{\rm bol, iso}$ (there is one model
having $m_{*} \sim 96\:M_{\odot}$) as our fiducial case. An early-type
star equivalent to an O6.5 star is postulated to have formed within
the HC HII region based on the derived Lyman continuum photon number
in Zhang et al. (2014). The molecular envelope shows evidence of
infall and outflow with an infall rate of $0.05 \ M_{\odot} \: \rm
yr^{-1}$ and a mass loss rate of $5.2 \times 10^{-3}\: M_{\odot} \: \rm
yr^{-1}$, which is somewhat higher than our derived disk accretion
rate, but may reflect infall on larger scales.
%They obtain rotational temperatures of $\sim$ 143$\pm$20 K for CH3CN(12-11), and $\sim$ 95$\pm$17 K for CH3CCH(13-12) and an electron temperature $T_{e}^{*} >$5500 K for the HC HII core.

\textbf{IRAS 16562-3959:}
There are only 4 fully valid data points constraining the
fitting. Since we have 5 free parameters and the $\chi^{2}$ is derived
by dividing the number of total data points including those as upper
limits, the small number of fully valid data points largely leads to
the relatively small $\chi^{2}$. The first four best models tend to
give high core masses $\sim$ 320 to 480 $M_{\odot}$ and low
$\Sigma_{\rm cl} \la 0.3 \: \rm g \: cm^{-2}$ clump environments, while
the fifth best model gives a less massive initial core of
$120\:M_{\odot}$ and a much denser $\Sigma_{\rm cl} \sim 3.2 \: \rm g
\: cm^{-2}$ clump. Note in the first three models the core radii are
larger than the aperture radius. The bolometric luminosity of the
source is reported to be $5-7\times 10^4\:L_{\odot}$ by Lopez et
al. (2011), which agrees well with most of the models. Guzman (2010)
also fit this source with Robitaille et al. (2007) models. The stellar
mass of their result 14.7 $M_{\odot}$ is close to our fourth and fifth
best models. Their disk accretion rate $5.5 \times 10^{-4} M_{\odot}
\ \rm yr^{-1}$ is closest to our fourth best model. Their envelope
mass 1700 $M_{\odot}$ is much larger than our results. Guzm\'an et
al. (2011) estimated the inclination angle of the SE-NW outflow to be
80$\arcdeg$, which is similar to our second best model.
%jctfinal - commenting out... outflow rate should be smaller than accretion rate.
%However, only before the correction for inclination is their mass
%outflow rate for the SE-NW outflow similar to the disk accretion rate
%returned by our models.

\textbf{G305.20+0.21:}
We have the most data for this source to constrain the model
fitting. The initial core mass returned is moderate, ranging from 80
to 160 $M_{\odot}$. Consistently, the envelope mass for this source is
also much lower than previous sources. The stellar mass ranges from 16
to 48 $M_{\odot}$, accreting at a high rate $\sim 10^{-3} M_{\odot} \:
{\rm yr}^{-1}$. Four models give $\Sigma_{\rm cl}$ as high as $3.2 \: \rm
g \: cm^{-2}$ and one gives $\Sigma_{\rm cl} \sim 1.0 \: \rm g
\: cm^{-2}$. The viewing angle is close to the outflow half opening
angle, resulting in a flat SED shape at short wavelengths. The
extrapolated IRAS luminosity is $\ga 10^5 \: L_{\odot}$ (Walsh et
al. 2001), which is consistent with the $L_{\rm bol}$ derived here.

\textbf{G49.27-0.34:}
The models at short wavelengths are much lower than the data points,
perhaps indicating significant PAH emission or small dust grain
emission from additional heating sources in the region. The best five
models all return $m_{*}$ of 12 $M_{\odot}$ and $\Sigma_{\rm cl}$ of
$3.2 \: \rm g \: cm^{-2}$. The initial core mass ranges from 120 to 320
$M_{\odot}$. The accretion rates are $\sim 10^{-3} \: M_{\odot} \: {\rm
  yr}^{-1}$.

\textbf{G339.88-1.26:}
The best four models prefer a protostar of $\sim 12 \: M_{\odot}$
accreting at $\sim 2 \times10^{-4} \: M_{\odot} \: \rm yr^{-1}$ in massive
cores of 320 to 480 $M_{\odot}$ in clumps with low $\Sigma_{\rm cl}
\sim 0.3 \: \rm g \: cm^{-2}$. Alternatively, the fifth best model gives a
less massive initial core mass of 120 $M_{\odot}$, but a much denser
clump environment with $\Sigma_{\rm cl} \sim 3.2 \: \rm g \: cm^{-2}$ and
a higher accretion rate of $\sim 10^{-3} \: M_{\odot} \: \rm yr^{-1}$. The
bolometric luminosity has been estimated to be 6.4$\times 10^{4}
L_{\odot}$ from the SED fitting to infrared fluxes with Robitaille et
al. (2007) models in Mottram et al. (2010, 2011), which is similar to
the luminosities in our five best models.

Recent ALMA observations (Zhang et al. 2019) reveal collimated
CO outflows with a half opening angle of $\sim$ 10$\arcdeg$. In
particular, they determine the outflow to be much edge-on so the
second model here with $i \approx 20\arcdeg$ is favored. They also
estimate the dynamical mass from the gas kinematics as $\sim$ 11
$M_{\odot}$, which is also consistent with our results.

\vspace{5mm}

In summary, the massive protostellar sources investigated in this
paper tend to have very massive initial cores, high protostellar
masses and high accretion rates. The mass surface densities of the
clump environments show significant variation.  The high envelope
masses indicate the protostars are still in an active stage of
accretion. Viewing angles tend to be more face-on than edge-on. This
allows shorter wavelength photons to more easily escape through the
outflow cavities towards the observer, though still regulated
partially by extinction of core infall envelope and foreground clump
material. Since SOMA survey sources have been selected based on their
previously known MIR emission, it is not surprising that the sample
may have such a bias towards having more face-on inclinations. Future
studies examining inclinations constrained from MIR image intensity
profiles and outflow kinematics will allow better measurement of
source orientations and a more thorough examination of this effect.

\section{Discussion}

\begin{figure}
%\figurenum{}
\epsscale{1.17}
\plotone{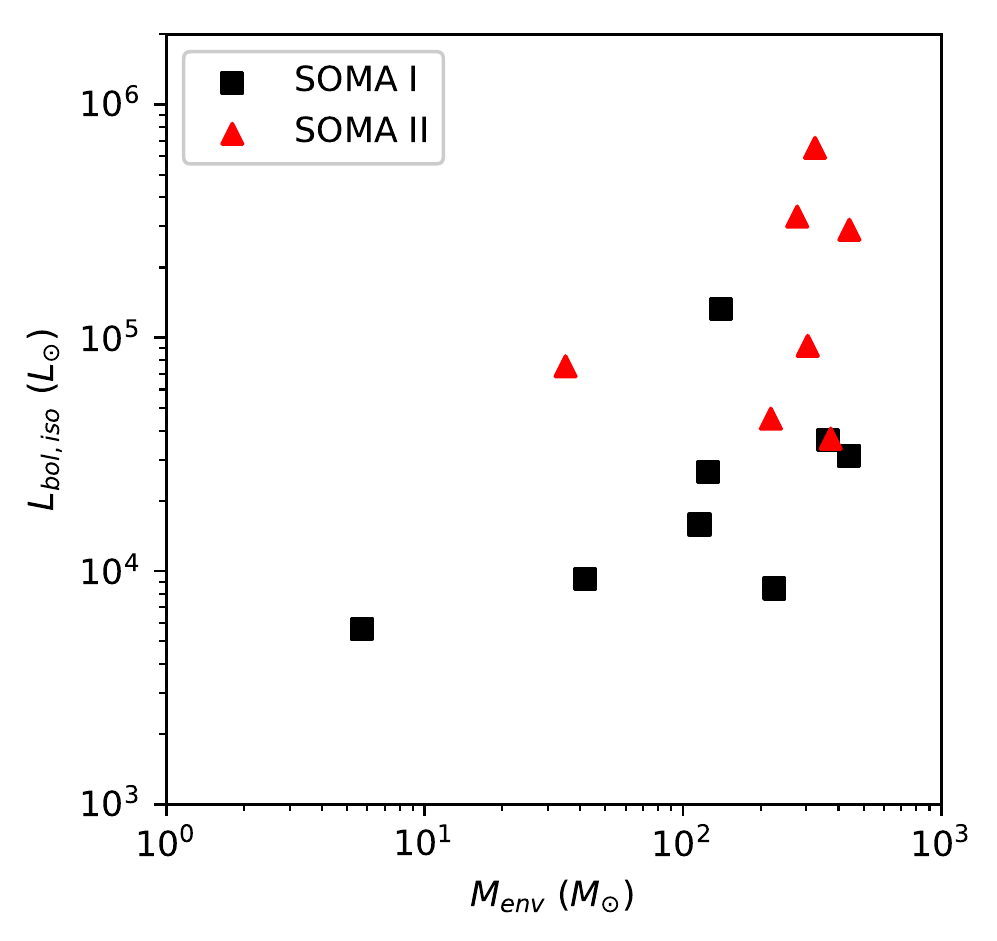}
\caption{
Diagram of isotropic luminosity versus the envelope mass returned by
the ZT best model. Squares denote the sample in Paper I. Triangles
denote the sample in this paper.}\label{fig:lm}
\end{figure}

\begin{figure*}
\epsscale{0.8}
\plotone{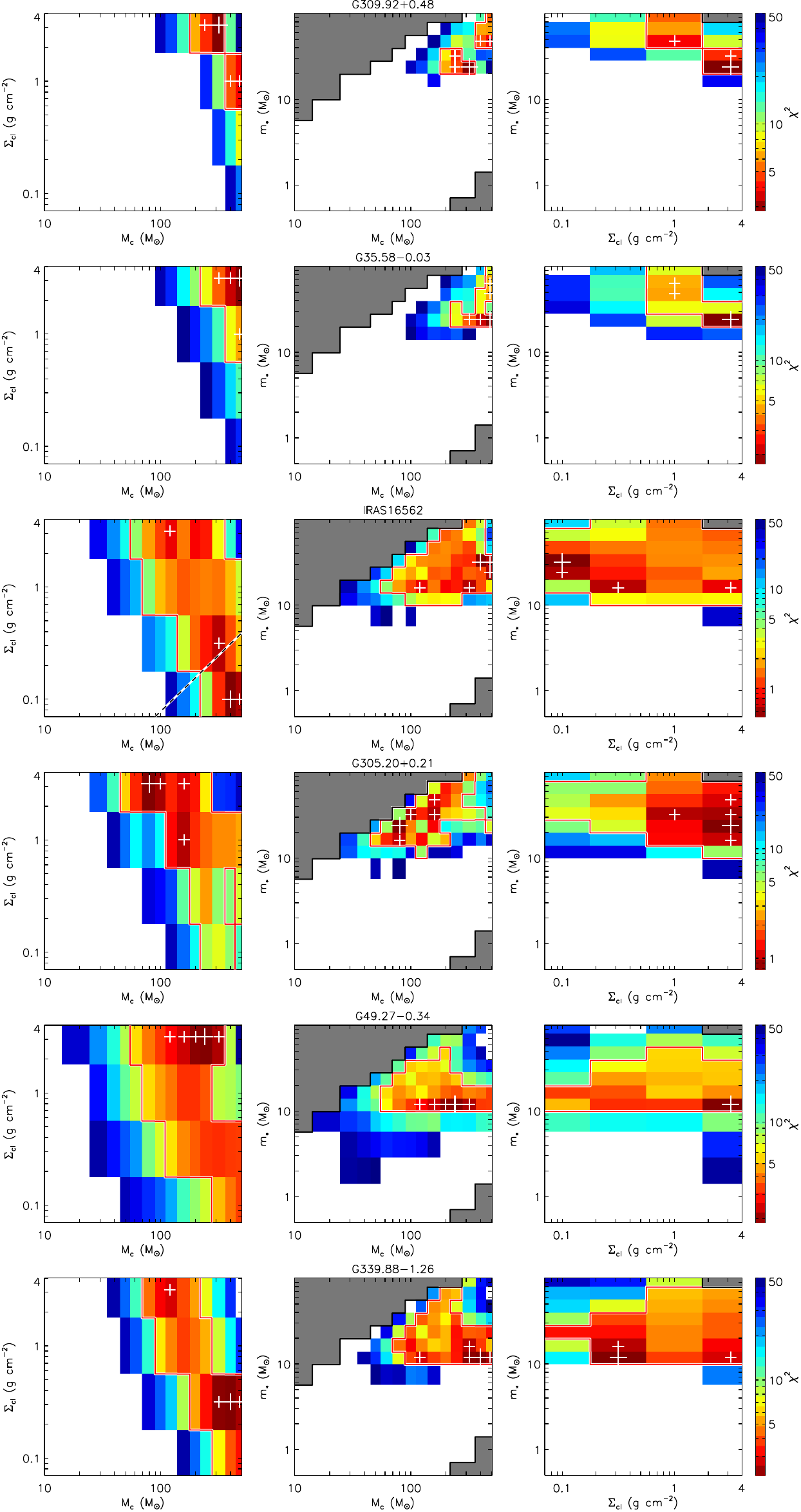}
\caption{
Diagrams of $\chi^{2}$ distribution in $\Sigma_{\rm cl}$ - $M_{c}$
space, $m_{*}$ - $M_{\rm c}$ space and $m_{*}$ - $\Sigma_{\rm
  cl}$ space. The white crosses mark the locations of the five best
models, and the large cross is the best model. The grey regions are
not covered by the model grid, and the white regions are where the
$\chi^{2}$ is larger than 50. The red contours are at the level of
$\chi^{2}$ = $\chi^{2}_{min}$ + 5. The dashed line denotes when $R_{c}
= R_{\rm ap}$.}\label{fig:primary}
\end{figure*}

\begin{figure}
\epsscale{1.15}
\plotone{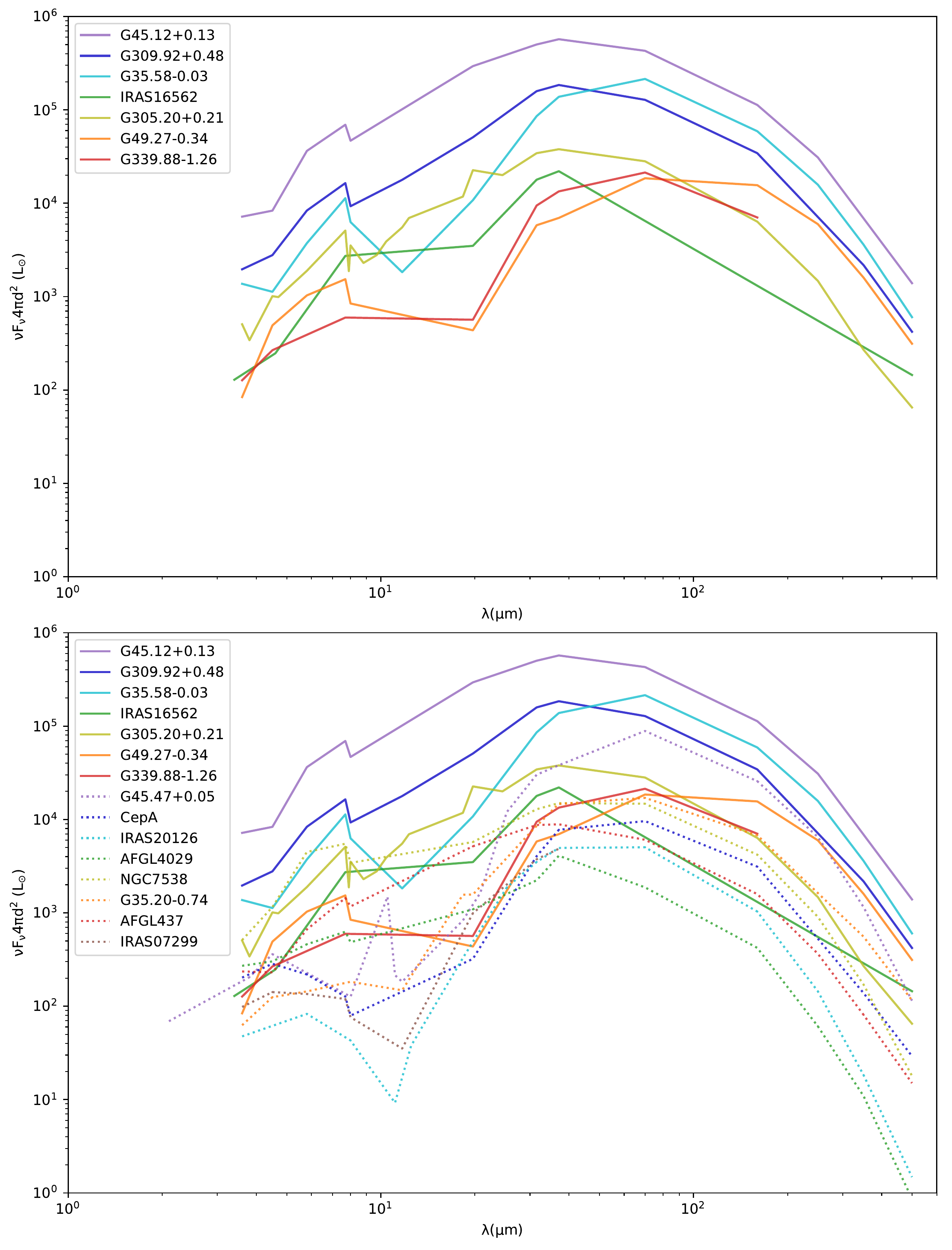}
\caption{
{\it Top panel:} Bolometric luminosity weighted SEDs of the eight SOMA
protostars analyzed in this paper. The ordering of the legend is from
high to low ZT best fit model isotropic luminosity (top to bottom). {\it Bottom
  panel:} Same as Top, but now with dotted lines denoting sample in
Paper I. }\label{fig:Lbol}
\end{figure}

\begin{figure}
%\figurenum{}
\epsscale{1.2}
\plotone{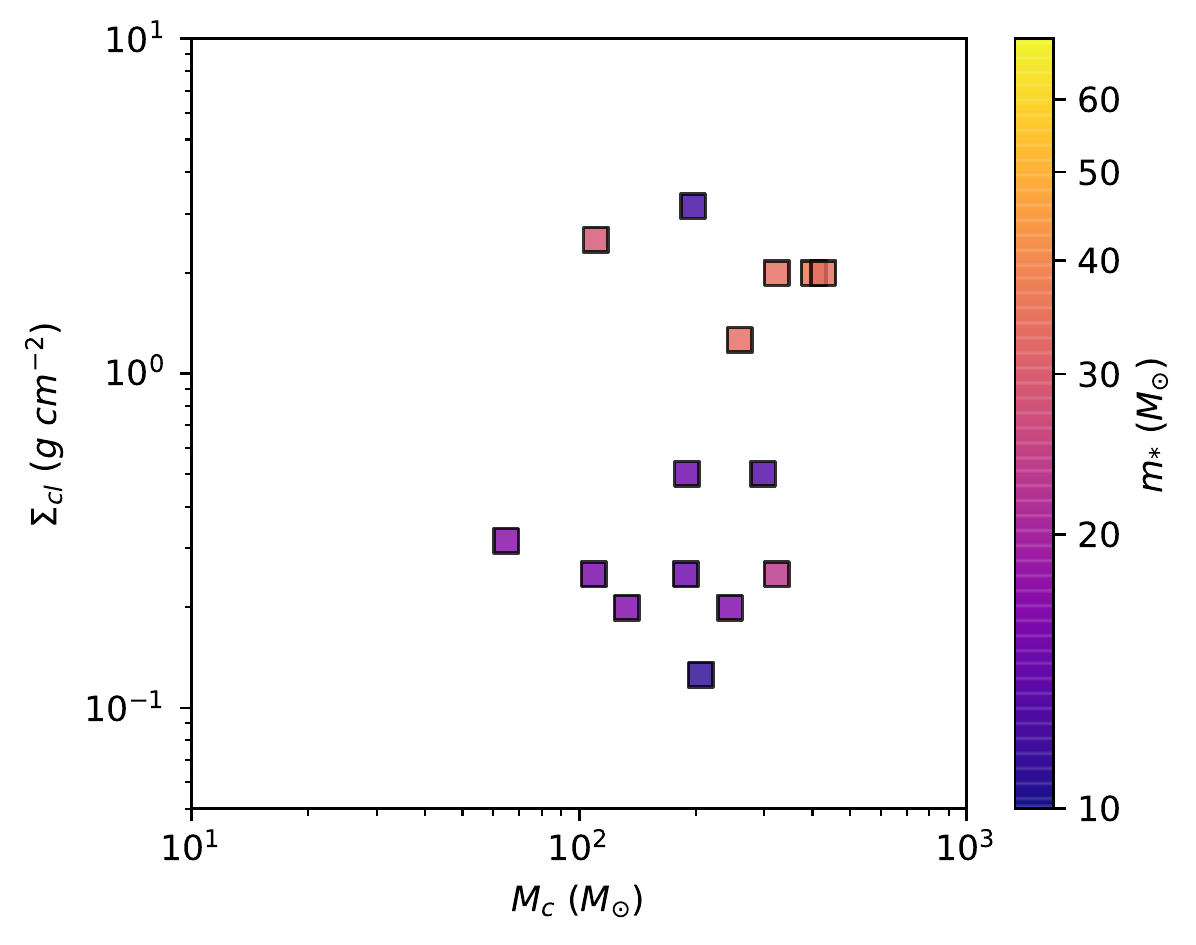}
\caption{
Diagram of the geometric mean clump surface density versus the
geometric mean initial core mass of the five best ZT models for each
source in Paper I and this work. The color indicates the geometric mean protostellar
mass.}\label{fig:m_sigma_ms}
\end{figure}

%jctfinal - commenting out this figure, see below. Maybe we can add back later after submission, but the trend does not look convincing.
%\begin{figure}
%%\figurenum{}
%\epsscale{1.2}
%\plotone{lSigma3.pdf}
%\caption{
%%
%Diagram of the geometric mean clump surface density versus the
%geometric mean isotropic luminosity of the five best ZT models for
%each source in Paper I and this work.  }\label{fig:lSigma}
%\end{figure}

Compared with the first eight protostars in Paper I, we have extended
the upper limit of the luminosity range by one order of magnitude as
shown in Figure~\ref{fig:lm}.
%and Figure~\ref{fig:Lbol}.
The seven sources in this paper are more luminous, and thus likely to
be more massive protostars embedded in higher mass cores. However,
there is the caveat of there being multiple sources sometimes present.

Figure~\ref{fig:primary} shows the $\chi^{2}$ distribution in
$\Sigma_{\rm cl}$ - $M_{c}$ space, $m_{*}$ - $M_{c}$ space and $m_{*}$
- $\Sigma_{\rm cl}$ space for 6 of the sources, i.e., all except
G45.12 due to its large $\chi^{2}$. These diagrams illustrate the full
constraints in the primary parameter space derived by fitting the SED
data, and the possible degeneracies among these parameters. Thus these
diagrams give a fuller picture of potential protostellar properties
than just the best five models.

Similar to Paper I, $M_{c}$ and $m_{*}$ are relatively well
constrained, while $\Sigma_{\rm cl}$ usually spans the full range (for
G49.27 the best five models return a universal $\Sigma_{\rm cl}$ of
3.2\,g\,cm$^{-2}$ though).
%This is partly because the grid in $\Sigma_{\rm cl}$ is rather sparse,
%with only four values.
The best models ($\chi^{2} - \chi^{2}_{\rm min} < 5$, within the red
contours) tend to occupy a region with lower $M_{c}$ at higher
$\Sigma_{\rm cl}$ and higher $M_{c}$ at lower $\Sigma_{\rm cl}$,
similar to the sources in Paper I as discussed in ZT18. The black
dashed line denotes a constant $R_{c}$ with $R_{c} = R_{\rm ap}$ using
$R_{c} = 0.057 (\Sigma_{\rm cl}/{\rm g \ cm}^{-2})^{-1/2}
\ (M_{c}/60M_{\odot})^{1/2}$ pc (MT03). Parameter sets higher than
this line mean they have a $R_{c}$ smaller than $R_{\rm ap}$, which is
more physical since we assume the aperture we choose covers the whole
envelope. This line only appears in IRAS 16562 because in other
sources the $R_{\rm ap}$ is so large that they all appear to the right
of the available $\Sigma_{\rm cl}$ - $M_{c}$ space. We can see for
most sources at least the best models satisfy this criterion.

In Figure~\ref{fig:Lbol} we show the bolometric luminosity spectral
energy distributions of the seven high luminosity protostars of this
paper, together with the sample from Paper I. Here the $\nu F_{\nu}$
SEDs have been scaled by $4 \pi d^{2}$ so that the height of the
curves gives an indication of the luminosity of the sources assuming
isotropic emission. The ordering of the vertical height of these
distributions is largely consistent with the rank ordering of the
predicted isotropic luminosity of the protostars from the best-fit ZT
models (the legend in Figure~\ref{fig:Lbol} lists the sources in order
of decreasing ZT best model isotropic luminosity). The curve of G305.20 appears
higher than IRAS 16562. However, if we look at all the five best models the
isotropic luminosity of G305.20 and IRAS 16562 are actually quite
close. The foreground extinction of G305.20 is also generally lower
than IRAS 16562, which leads to a higher $4 \pi d^{2} \nu
F_{\nu}$. Similarly, the foreground extinction of G339.88 is on
average lower than G49.27, so that G339.88 has a larger height of the
bolometric luminosity SED.

We find no obvious systematic variation in SED shape with varying
luminosity. This was investigated by plotting the slope between
19\,$\mu$m and 37\,$\mu$m versus the isotropic luminosity of the
sources (not shown here). We also investigated the relation between
$\Sigma_{\rm cl}$, $M_{c}$ and $m_{*}$ in
Figure~\ref{fig:m_sigma_ms}. To form high-mass stars naturally
requires relatively massive cores (this assumption is built in to the
models).
%there may be a threshold in $M_{c}$ but
However, $\Sigma_{\rm cl}$ does not have to be very high. However, the
models with $\Sigma_{\rm cl} \sim 0.1 \: \rm g\: cm^{-2}$ have $R_{\rm
  core} > R_{\rm ap}$ most of the time, which is physically
inconsistent with the analysis method. The models with $\Sigma_{\rm
  cl} \sim 0.3 \: \rm g\ cm^{-2}$ only have $R_{\rm core} > R_{\rm
  ap}$ occasionally, while the other models with higher $\Sigma_{\rm
  cl}$ do not have such a problem. Thus it is massive protostellar
core models with $\Sigma_{\rm cl} \ga 0.3 \: \rm g\: cm^{-2}$
surrounding clump environments that are currently consistent with the
observed sources.

Overall the ZT models can fit the observed SEDs reasonably well
assuming a single protostar forming through an axisymmetric monolithic
collapse from a massive core. Only in G45.12, which has stronger
evidence for their being multiple protostars that are part of a
forming cluster, do the models fare badly and have relatively large
values of $\chi^{2}$ (although this may also be due to its extreme
luminosity causing it to be near the edge of the ZT model grid). There
are reported examples of quite ordered protostellar cores, i.e., with
collimated, symmetric outflows: e.g., the case of the early-stage
protostar C1-Sa (Tan et al. 2016) and G339.88-1.26 (Zhang et al. 2019,
presenting follow-up ALMA observations of one of these SOMA
sources). On the other hand, there are also cases that appear much
more disordered in both their accretion flows (W51e2e, W51e8, and W51
north, Goddi et al. 2018) and outflows (Orion KL, Bally et
al. 2017). The combination of MIR to FIR SED and image fitting with
high resolution studies of infall and outflow morphologies for larger
samples will allow us to better determine the limitations of simple
axisymmetric protostellar core models for Galactic massive star
formation studies.

\section{Conclusions}

We have presented the results of MIR and FIR observations made towards
the next seven highest luminosity protostars in the SOMA survey, built
their SEDs and fit them with RT models of massive star formation via
the Turbulent Core Accretion model. Our goal has been to expand the
observational massive protostar sample size to test the star formation
models over a wider range of properties and environments and
investigate trends and conditions in their formation. Compared with
the first eight protostars in Paper I, the seven YSOs in this paper
are more luminous, and thus likely to be more massive protostars. Some
of the new sources appear to be in more clustered environments and/or
have lower-mass companions relatively nearby. In summary, our main
results and conclusions are as follows.

1. The MIR emission of massive protostars is strongly influenced by
outflow cavities, where extinction is relatively low. We see MIR
extension along detected outflows in IRAS16562 and G339.88. Away from
these cavities, extinction can be very high and block MIR emission.
There is also a hint that the MIR emission may reveal the presence of
the optically thick disk perpendicular to the outflow as in G309.92
and G305.20, though more evidence of the position of the protostar
from mm or radio continuum observations will be needed to confirm the
disk. The high extinction in the MIR tells us that large quantities of
high column density material is present close to the protostar, as
expected in the Turbulent Core model.

2. The sources span a luminosity range of $10^{4} -10^{6} \:
L_{\odot}$. Fitting the SEDs with RT models yields protostellar masses
$m_{*} \sim 12-64 \: M_{\odot}$ accreting at rates of $\dot{m}_{*}
\sim 10^{-4}-10^{-3} \: M_{\odot} \: \rm yr^{-1}$ inside cores of
initial masses $M_{c} \sim 100-500 \: M_{\odot}$ embedded in clumps
with mass surface densities $\Sigma_{\rm cl} \sim 0.1-3 \: \rm g
\ cm^{-2}$. The relatively high protostellar mass in several sources
is possibly due to there being more than one protostar in the region
and the $m_{*}$ derived could be the sum of multiple sources.

3. The SED shape, especially the slope at short wavelengths, appears
related to the viewing angle and the outflow opening angle. When the
viewing angle is close to the outflow opening angle, a relatively flat
slope at short wavelengths results. However, the SED shape, especially
the location of the SED peak, is also likely to be related to the
evolutionary stage of the protostar: more evolved protostars tend to
peak at relatively shorter wavelengths. So far we do not see obvious
relations between SED shape and bolometric luminosity.

4. To form high-mass stars naturally requires high values of $M_{c}$,
but not seem to require especially high values of $\Sigma_{\rm cl}$.
%does not have to be high.
We see high-mass protostars are able to at least form from
$\Sigma_{\rm cl} \ \ga \ 0.3 \ \rm g\ cm^{-2}$ environments.
%However, there is a slight hint that more luminous sources tend to be
%associated with a higher $\Sigma_{\rm cl}$ environment. There is also
%a slight hint that more massive and evolved protostellar sources
%(indicated by $m_{*}$) tend to be found in environments with
%relatively high $M_{c}$ and $\Sigma_{\rm cl}$.

5. Radiative transfer models based on the Turbulent Core Accretion
scenario can reasonably well describe the observed SEDs of most
relatively isolated massive protostars, but may not be valid for the
most luminous regions in the sample, which may be better treated as
protoclusters containing multiple sources. Whether or not core
accretion models can apply on smaller physical scales within these
regions requires higher angular resolution MIR to FIR observations.

%High mass surface density, high pressure regions are more likely to
%harbor high-mass protostars, but the high density does not seem
%necessary.
%jct - inconsistent...

%jct 

\acknowledgments We thank the anonymous referee for reading the pa- per carefully and providing useful comments. We thank Crystal Brogan and Adam Ginsburg for helpful discussion and suggestions. 
M.L. and J.C.T. acknowledge funding from NASA/USRA/SOFIA. J.C.T. acknowledges support from NSF grant AST1411527 and ERC project 788829 - MSTAR. K.E.I.T. acknowledges support by NAOJ ALMA Scientific Research Grant Numbers 2017-05A. This work is additionally based on observations
obtained at the Gemini Observatory [programs GS-2004A-Q-7 and
GS-2005A-DD-5], which is operated by the Association of Universities
for Research in Astronomy, Inc., under a cooperative agreement with
the NSF on behalf of the Gemini partnership: the National Science
Foundation (United States), the National Research Council (Canada),
CONICYT (Chile), Ministerio de Ciencia, Tecnologa e Innovacin
Productiva (Argentina), and Ministrio da Cincia, Tecnologia e Inovao
(Brazil).

%jctfinal - commenting these out... maybe can bring them in later, or in Paper III.

%\begin{figure}
%%\figurenum{}
%\epsscale{1.17}
%\plotone{flat_Liso.pdf}
%\caption{slope between 19$\mu$m and 37$\mu$m vs. the isotropic luminosity. Blue dots denote the luminosity returned by the best model. Orange dots denote the geometric mean of the luminosities of the best 5 models.
%}\label{fig:flat}
%\end{figure}

%\begin{figure}
%%\figurenum{}
%\epsscale{1.2}
%\plotone{m_sigma_ms2.pdf}
%\caption{
%%
%Diagrams of the clump surface density versus the initial core mass
%of the five best ZT models for each source in Paper I and this work. The color indicates the
%stellar mass. For better illustration, the clump surface density of each data point plotted is randomly scaled by a factor within 1.2 to avoid overlap.}
%\end{figure}

%\begin{figure}
%%\figurenum{}
%\epsscale{1.2}
%\plotone{lSigma2.pdf}
%\caption{Diagrams of the clump surface density versus the isotropic luminosity of the five best ZT models for each source in Paper I and this work. For better illustration, the clump surface density of each data point plotted is randomly scaled by a factor within 1.2 to avoid overlap.
%}
%\end{figure}

\end{document}